\documentclass[number,sort&compress,5p]{elsarticle}
\pdfoutput=1
\usepackage{amsmath,amssymb}
\usepackage{mathptmx}
\usepackage{graphicx}
\usepackage{hyperref}
\usepackage{longtable}
\usepackage{xcolor}
\usepackage{multirow}
\usepackage{lscape}
\usepackage{subfigure}
\usepackage{epsfig}
\usepackage{supertabular}
\usepackage{epstopdf}
\usepackage{amssymb}
\usepackage{bm}
\usepackage{etoolbox}
\usepackage{tablefootnote}
\usepackage[flushleft]{threeparttable}
\DeclareMathOperator\erf{erf}
\newcommand{\appendices}{
  \renewcommand{\thesection}{\Alph{section}}
  \setcounter{section}{0}
  \renewcommand{\theequation}{\Alph{section}-\arabic{equation}}
  \setcounter{equation}{0}
  \renewcommand{\thefigure}{\Alph{section}-\arabic{figure}}
  \setcounter{figure}{0}
  \renewcommand{\thetable}{\Alph{section}-\arabic{table}}
  \setcounter{table}{0}
}

\newcommand*\aap{A\&A}

\newcommand*\aj{AJ}

\newcommand*\apj{ApJ}
\newcommand*\apjl{ApJ}

\newcommand*\apjs{ApJS}

\newcommand*\apss{Ap\&SS}

\newcommand*\jcap{J. Cosmology Astropart. Phys.}

\newcommand*\mnras{MNRAS}
\newcommand*\na{New A}

\newcommand*\nat{Nature}

\newcommand*\pasp{PASP}

\newcommand*\prd{Phys.~Rev.~D}

\begin{document}
\title{On the absence of a universal surface density, and a maximum Newtonian acceleration in dark matter haloes: consequences for MOND}

\author[IHEP,UCAS]{Yong Zhou}
\author[Catania,INFN,INASAN]{A. Del Popolo}
\ead{adelpopolo@oact.inaf.it}
\author[IHEP,UCAS]{Zhe Chang}

\address[IHEP]{Institute of High Energy Physics, Chinese Academy of Sciences, Beijing 100049, China}

\address[UCAS]{School of Physical Sciences, University of Chinese Academy of Sciences, Beijing 100049, China}

\address[Catania]{Dipartimento di Fisica e Astronomia, University Of Catania, Via S. Sofia, 64, 95123 Catania, Italy}

\address[INFN]{INFN Via S. Sofia, 64, 95125 Catania CT}

\address[INASAN]{Institute of Astronomy of the Russian Academy of Sciences, Pyatnitskaya  Str. 48, 109017 Moscow, Russia}

\date{}

\begin{abstract}\noindent We study the dark matter (DM) surface density using the SPARC sample and {compare} it to Donato et al. \cite{Donato} result. By means of MCMC method, we infer the best-fitting parameters for each galaxy. We reobtain the scaling relation between the surface density and luminosity, and several other scaling laws relating the dark matter halo properties to that of the galactic disc properties. We conclude, in contrast with Donato et al. \cite{Donato}, that the dark matter surface density is not a universal (constant) quantity but correlates with the luminosity as well as with other galactic disc properties. A derived posterior probability distribution of $\rho_0 r_0$ shows that the null hypothesis of constancy is rejected at a very high confidence level. These results leave little room for the claimed universality of dark matter surface density. Since MOND has strong prediction on the surface density \cite{Milgrom2009}, we compared our result with those predictions, finding that MOND predictions are violated by data. To strengthen the previous result, we compared our results to another prediction of MOND \cite{Milgrom2005}, the existence of a maximum Newtonian dark matter acceleration in the halo. Also in this case, MOND predictions are in contradiction with data. The dark matter Newtonian acceleration correlates with all the previously presented galactic disc properties, and data are distributed outside the bound predicted by Milgrom $\&$ Sanders\cite{Milgrom2005}. We also find that the null hypothesis (constancy of DM Newtonian acceleration) is rejected at a very high confidence level.
\end{abstract}

\maketitle

\section{Introduction}

The $\Lambda$ cold dark matter ($\Lambda$CDM) model, or concordance cosmology, gives very accurate predictions of the observations on cosmological scales\footnote{We recall that at this scales the cosmological constant problem \cite{Weinberg, Astashenok}, and the cosmic coincidence problem affect the $\Lambda$CDM paradigm.} \cite{2011ApJS..192...18K,DelPopolo2007,2013AIPC15482D}, and intermediate scales \cite{Spergel,Kowalski,Percival,2011ApJS..192...18K,DelPopolo2007,2013AIPC15482D,2014IJMPD2330005D}. 

For precision's sake, even at large scale there are tensions of unknown origin between the value of the Hubble parameter, $H_0$, and SNe Ia data, the 2013 Planck parameters \cite{planck} and $\sigma_8$ obtained from cluster number counts and weak lensing. Also the Planck 2015 data are in tension with  $\sigma_8$ growth rate \cite{maca}, and with CFHTLenS weak lensing data \cite{raveri}. Moreover, 
a quadrupole-octupole alignment \cite{schwa,copi,copi1,copi2,copi3}, a power hemispherical asymmetry \cite{eriksen,hansen,jaf,hot,planck1,akrami} and a cold spot \cite{cruz,cruz1,cruz2} are presented in the 
large-angle fluctuations in the CMB. Moving to smaller scales ($\simeq 1-10$ kpcs), the $\Lambda$CDM model is affected by a series of problems \cite{moore94,moore1,ostrik,boyl,boyl1,oh,DelPopoloHiotelis2014,DelPopolozz,DelPopoloPace2016,DelPopolo2017,Zhou:2017lwy,Chang:2018vxs,2019MNRAS.486.1658C}.
Some of them are
\renewcommand{\theenumi}{\textbf{\alph{enumi}}}\begin{enumerate} 
	\item the cusp/core (CC) problem \cite{moore94,flores}, that is the discrepancy in the inner profiles obtained in dissipationless N-body simulations \cite{nfw1996,nfw1997,navarro10} and observations 
	of dwarf galaxies, Irregulars, and Low Surface Brightness {(LSB)} galaxies \cite{deBloketal2003, Swatersetal2003,DelPopolo2009,DelPopoloKroupa2009,2011AJ141193O,KuziodeNaray2011,2012MNRAS419971D,2012MNRAS42438D,DelPopoloHiotelis2014}, and clusters \cite{DelPopolo2014z};
	\item the ``missing satellite problem"
	(MSP), namely the discrepancy between the number of sub-haloes predicted by N-body simulations \cite{moore1}, and observations;
	\item the ``Too-Big-To-Fail problem", namely the fact that sub-haloes are too dense compared to what we observe around the Milky Way \cite{GarrisonKimmel2013,GarrisonKimmel2014};
	\item the satellites planes problem, that is the difficulty in explaining the location of satellite galaxies of the Milky Way and M31 on planes \cite{pawl}.
\end{enumerate}

Between the solutions proposed, we recall the proposal of modifying the theory of gravity \cite{1970MNRAS1501B,1980PhLB9199S,1983ApJ270365M,1983ApJ270371M,Ferraro2012,Rodrigues:2018duc}, modifying the nature of DM \cite{2000ApJ542622C,2000NewA5103G,2000ApJ534L127P,2001ApJ551608S}, modifying the power spectrum \cite{2003ApJ59849Z}, or delegating the solution to the phenomena related to baryon physics phenomena, which are complex and not well understood.  Two of the mechanisms proposed are related to supernovae explosions \cite{1996MNRAS283L72N,Gelato1999,Read2005,Mashchenko2006,Governato2010}, or to transfer of energy and angular momentum from baryons to DM through dynamical friction \cite{El-Zant2001,El-Zant2004,2008ApJ685L105R,DelPopolo2009,Cole2011,2012MNRAS419971D,Saburova2014,DelPopoloHiotelis2014}.

In order to understand complex phenomena, scaling relations are very helpful.

Kormendy $\&$ Freeman \cite{Kormendy2004} found several {scaling} relations modeling the rotation curves of galaxies through a pseudo-isothermal (pISO) profile. They found several relations between DM halos parameters. 

One of those relations, namely $\rho_0 r_c$, where $\rho_0$ is the central density of the density profile, and $r_c$ is its scale radius, was shown to be independent, in the case of late-type galaxies, from galaxy luminosity. Kormendy $\&$ Freeman \cite{Kormendy2004} found that $\rho_0 r_c \simeq 100 {\rm \,M_{\odot}\,pc^{-2}}$.

Several other authors went on studying the quoted relation, and in the case of Donato et al. \cite{Donato} (hereafter D09), the 55 galaxy sample of Kormendy $\&$ Freeman \cite{Kormendy2004} was extended by means of $\simeq 1000$ spiral galaxies, 
weak lensing of spirals and ellipticals, and data from dwarf galaxies. They found a similar result to that of 
Kormendy $\&$ Freeman \cite{Kormendy2004}, which led them to claim a quasi-universality of 
$\Sigma^0_{\rm Donato}= \rho_0 r_0$\footnote{$\rho_0$ and $r_0$ are the {central} density and scale radius of the Burkert profile.},
interpreted as central surface density of DM halos. Shortly after the publication of the D09 paper, Milgrom \cite{Milgrom2009} showed that in the Newtonian regime, the modified Newtonian dynamics (MOND) paradigm predicted very similar results to that of D09. 

A  further extension of the D09 result was that of Gentile et al. \cite{Gentile2009} (hereafter G09), claiming a quasi-universality of the luminous surface density within scale radius of the dark halo.

A common feature to D09, and G09 is that they used the same sample, and  assumed that a very different class of galaxies, going from dwarfs to ellipticals, could be fitted by the same halo density profile, namely the Burkert profile
\begin{equation}
\rho(r)=\frac{\rho_0 r_0^3}{(r+r_0)(r^2+r_0^2)}
\end{equation}
obtaining $\rho_0$, and $r_0$. 
Now, the Burkert profile is known to give good fits to dwarf galaxies, and LSBs, but not to elliptical galaxies. The DM distribution obtained with different methods (X-ray properties of the emitting hot gas
\cite{Nagino2009,Buote2012}, stellar dynamics \cite{Gerhard2010,Napolitano2011}, weak and strong lensing \cite{An2013,Lyskova2018}), can be fitted both by the NFW profile or the isothermal profile (a cored profile like the Burkert profile), at least for X-ray data. de Blok et al. \cite{THINGS} found that brighter, larger galaxies with $M_{\rm B} > -19$ have density profiles well fitted by both cuspy profiles and cored ones, while less massive galaxies with $M_{\rm B} < -19$ are best fitted by cored profiles.

Moreover, even in the case of dwarfs, Simon et al. \cite{Simon2005} showed that cored profiles, like Burkert profile, are not a good fit to some of them, and some dSphs could have a cuspy profile \cite{Strigari2010,Breddels2013}, instead of a cored one, as expected. 

The quoted results bring us at least to have some doubts about the D09, and G09 conclusions, since they are based on the assumption that all the galaxies they studied are well fitted by {the} Burkert profile, and then they are all cored. 

Opposite results to that of D09, and G09 were obtained by Napolitano et al. \cite{Napolitano2010} who showed that the projected density within the local radius is larger in the case of early type galaxies with respect to that of dwarfs and spirals. Boyarsky et al. \cite{Boyarsky} arrived to similar conclusions with a sample containing group of galaxies, and clusters, much larger than the one of D09, and G09.

They showed that the projected density of local early type galaxies, within the effective radius is larger than that of dwarfs and spirals. This systematic increase with the mass of the halo was also noticed by Boyarsky et al. \cite{Boyarsky}. The  sample of Boyarsky et al. \cite{Boyarsky} was larger than that of D09, and G09, including groups and clusters. The dark matter column density, $S$, defined by them,  is increasing with the halo mass as
\begin{eqnarray}
\log_{10} S= 0.21\,\log_{10} \frac{M_{\rm halo}}{10^{10}\, {\rm M_{\odot}}}+1.79
\end{eqnarray}
with $S$ in ${\rm M_{\odot}\,pc^{-2}}$. 
Also Cardone $\&$ Tortora \cite{CardoneTortora2010} showed that the column density and the Newtonian acceleration, obtained through strong lensing, and central velocity dispersion of local galaxies are not constant but correlates, in agreement with Boyarsky et al. \cite{Boyarsky}, with the halo mass $M_{200}$, also with the stellar mass $M_{\ast}$, and the visual luminosity. Napolitano et al. \cite{Napolitano2010} 
found that the constant density scenario is violated by  the early-type galaxies. 
A correlation of the surface density with $M_{200}$, and between the baryon column density and mass, 
was obtained by Del Popolo et al. \cite{2013MNRAS4291080D}. {Different} from \cite{Napolitano2010}, \cite{Boyarsky} and \cite{CardoneTortora2010}, the result had a smaller scatter. Cardone $\&$ Del Popolo \cite{CardoneDelPopolo2012} followed closely D09, and G09 analysis, doubling the sample of the previous authors, also investigating selection effects, and reobtained the halo parameters fitting all the galaxy, while D09, and G09, for many galaxies obtained from literature. Again they found a result in contradiction to that of D09, and G09, since Newtonian acceleration and virial mass were found to be correlated. 

Finally, a more recent result by Li et al. \cite{Li:2018rnd}, found that fitting the haloes with an Einasto profile, and a DC14 \cite{DiCintio2014} profile the surface density is correlated with luminosity. 

We recall that the D09, and especially G09 paper were clearly hinting to a relation of their results with MOND predictions, and was a confirmation of those predictions. This was already confirmed by Milgrom \cite{Milgrom2009}. The results of the previously quoted papers \cite{Boyarsky,Napolitano2010,CardoneTortora2010,CardoneDelPopolo2012,2013MNRAS4291080D,Saburova2014,Li:2018rnd}
showing that the surface density is not a universal quantity, in strong contradiction to D09, and G09, is simply saying that the surface density result contradicts MOND expectation. We will discuss this point in this paper, reobtaining the surface density with the same profile used by D09, and G09, and inferring the parameters through the Markov chain Monte Carlo (MCMC) method. We will also discuss another MOND prediction given by Milgrom $\&$
Sanders \cite{Milgrom2005} concerning the existence of a maximum halo acceleration, and using similar methods we show again that data {contradicts} MOND predictions.

The paper is organized as follows. In Sec. \ref{Methodology}, we describe the methodology. In Sec. \ref{Result}, we analyze the MCMC results and the scaling relations between the dark halo and the galactic disc properties. In Sec. \ref{MOND}, we discuss the impact of our results on MOND. Discussion and conclusions are given in Sec. \ref{Conclusions}.

\section{Methodology}
\label{Methodology}

\subsection{SPARC data set}
\label{Data}

The \textit{Spitzer} Photometry and Accurate Rotation Curves (SPARC) data set \footnote{\url{http://astroweb.cwru.edu/SPARC/}} \cite{Lelli:2016zqa} is a sample of 175 late-type disc galaxies with new surface photometry at 3.6 $\mu$m and high-quality rotation curves from previous HI/H$\alpha$ studies. The surface photometry at 3.6 $\mu$m provides the stellar mass via the mass-to-light ratio $\Upsilon_*$ conversion factor. In the near infrared bands, $\Upsilon_*$ has small changes with star formation history  
\cite{McGaugh2014,Meidt2014}, and the distribution of the stellar mass are well determined by \textit{Spitzer} photometry. The 21cm observations provide the gas mass. The majority of SPARC galaxies are characterized by a disc structure, and some have bulges, both of them constitute the stellar component.  In total, the galaxy baryonic mass profile includes disc, bulge and gas component and the dark matter profile will be introduced later. In the SPARC data set, the mass profile is represented by velocity at a given radius, so the total baryonic velocity is
\begin{eqnarray}\label{eq:vbar}
V_{\rm bar}^2=\Upsilon_{\rm d}V^2_{\rm disc}+\Upsilon_{\rm b}V^2_{\rm bulge}+V^2_{\rm gas},
\end{eqnarray}
where $\Upsilon_{\rm d}$ and $\Upsilon_{\rm b}$ are the mass-to-light ratios for disc and bulge component, respectively.
SPARC spans a wide range of morphologies (S0 to Irr), luminosities (5 dex), and surface brightnesses (4 dex). SPARC sample is particularly good to study both DM haloes, and the way they are related to the discs of galaxies.

\subsection{Dark halo profile}
\label{DarkHalo}

In order to compare our results to that of D09, and G09, in this paper we will use {the} Burkert profile 
dark halo to fit the SPARC data set. The choice of this profile is dictated by the fact D09, and G09 used the same profile in their analysis. 
%
%
The Burkert density profile is given by
\begin{eqnarray}
\rho(r)=\frac{\rho_0r_0^3}{(r+r_0)(r^2+r_0^2)},
\end{eqnarray}
where $\rho_0$ and $r_0$ are the central density and scale radius of a halo, respectively.
Its enclosed mass profile is given by
\begin{eqnarray}
M(r)=2\pi\rho_0 r_0^3 [\ln{(1+x)}+\frac{1}{2}\ln{(1+x^2)}-\arctan{x}],
\end{eqnarray}
where $x=r/r_0$ is a dimensionless radius.
The rotation velocity from DM haloes is given by
\begin{eqnarray}
\frac{V_{\rm DM}^2}{V_{200}^2}=\frac{C_{200}}{x}\frac{\ln{(1+x)}+\frac{1}{2}\ln{(1+x^2)}-\arctan{x}}{\ln{(1+C_{200})}+\frac{1}{2}\ln{(1+C_{200}^2)}-\arctan{C_{200}}}.
\end{eqnarray}
The concentration $C_{200}$ and the rotation velocity $V_{200}$ at the virial radius $r_{200}$ are given by
\begin{eqnarray}
\label{eq1}
C_{200}=r_{200}/r_0,~~V_{200}=10C_{200}r_0H_0,
\end{eqnarray}
where $H_0$ is the Hubble constant ($73~\mathrm{Km\,s^{-1}\,Mpc^{-1}}$ in this paper).

The total rotational velocity is given by summing all the components, as
\begin{eqnarray}
V_{\rm tot}^2=V_{\rm DM}^2+\Upsilon_{\rm d}V^2_{\rm disc}+\Upsilon_{\rm b}V^2_{\rm bulge}+V^2_{\rm gas},
\end{eqnarray}
where $V_{\rm DM}$ is the dark matter component, and $V_{\rm disc},~V_{\rm bulge},~V_{\rm gas}$ the baryonic component, respectively. 
$\Upsilon_{\rm d}$, and $\Upsilon_{\rm b}$ 
represent the mass-to-light ratios for the disc and bulge component, which is predicted by the stellar population synthesis model \cite{McGaugh2014,Meidt2014}. 
Apart from the stellar mass-to-light ratios, the galaxy distance and disc inclination affect the stellar components and the total observed rotational velocities, respectively. If the galaxy distance $D$ is changed to $D'=D\delta_{\rm D}$, where $\delta_{\rm D}$ is a dimensionless distance factor, then the radius changes according to $R'=R\delta_{\rm D}$ and the baryonic component velocity changes to $V'_{\rm k}=V_{\rm k}\sqrt{\delta_D}$, where `k' denotes disc, bulge, or gas. If the disc inclination is changed to $i'=i\delta_{\rm i}$, where $\delta_{\rm i}$ is a dimensionless inclination factor, the observed rotation curves and its uncertainties change according to
\begin{eqnarray}
V_{\mathrm{obs}}^{\prime}=V_{\mathrm{obs}} \frac{\sin (i)}{\sin \left(i^{\prime}\right)}, \quad \delta V_{\mathrm{obs}}^{\prime}=\delta V_{\mathrm{obs}} \frac{\sin (i)}{\sin \left(i^{\prime}\right)}.
\end{eqnarray}
Then, we can compare the total rotational velocity with the observed rotation velocity. In total, the free parameters in the fits we will perform are: $V_{200}$, $C_{200}$, $\Upsilon_{\rm d}$, $\Upsilon_{\rm b}$, $\delta_{\rm D}$ and $\delta_{\rm i}$.

\begin{figure*}
	\begin{center}
		\includegraphics[width=0.33\textwidth]{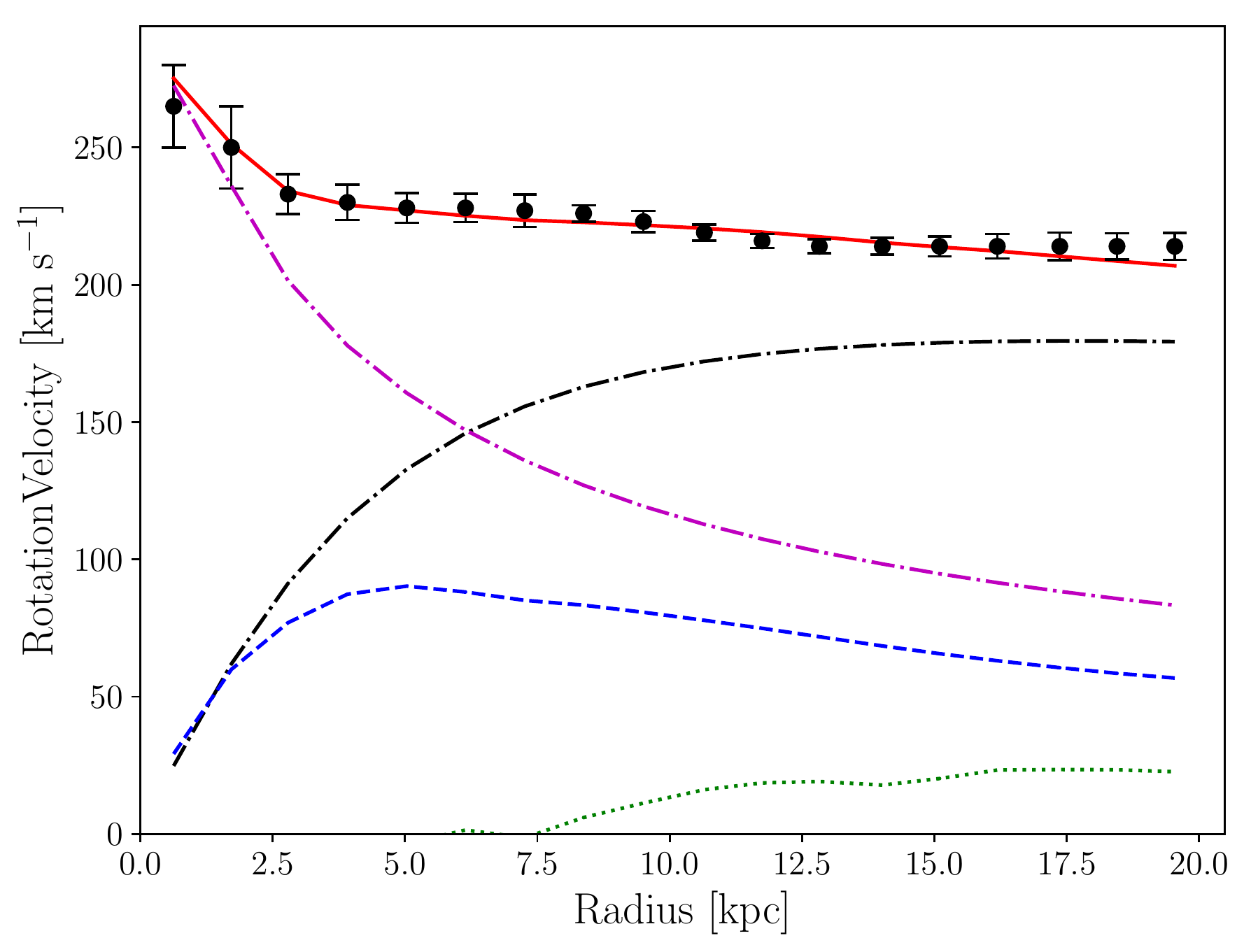}
		\includegraphics[width=0.33\textwidth]{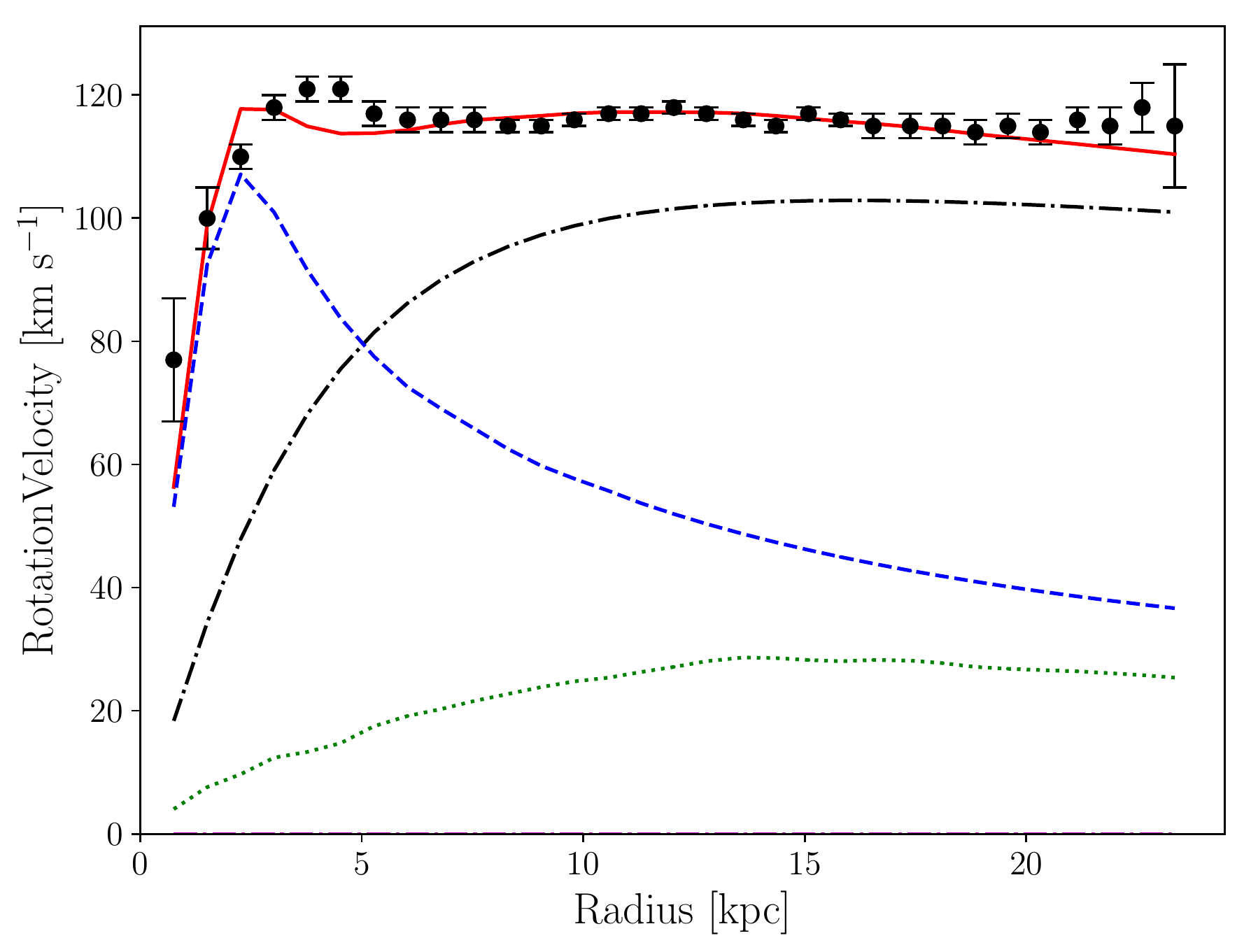}
		\includegraphics[width=0.33\textwidth]{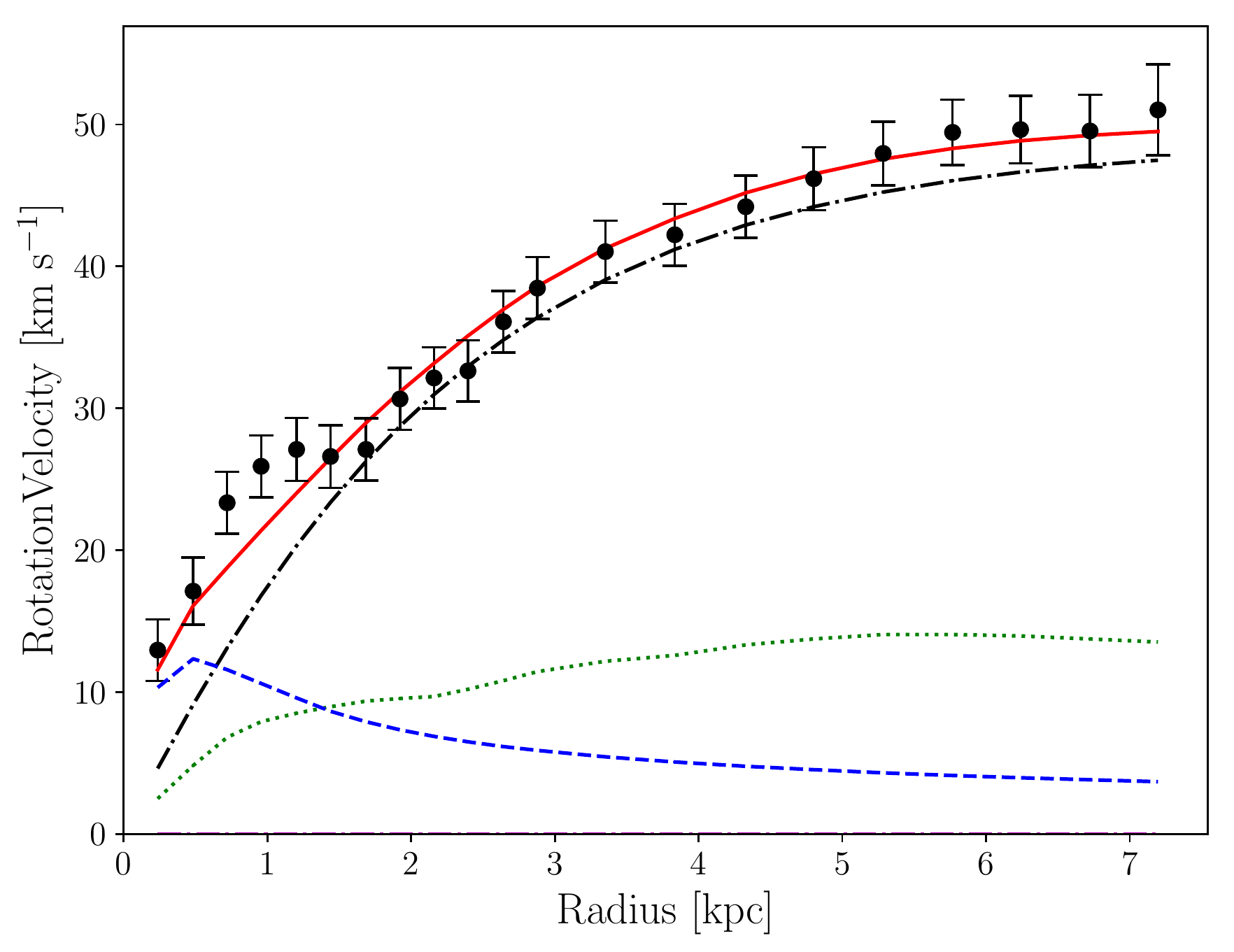}
		\caption{The best fit of galaxy rotation curve for the bulge-dominated spiral galaxy (NGC 7814, left), disc-dominated spiral galaxy (NGC 6503, middle) and gas-dominated dwarf galaxy (NGC 3741, right). The points with error bars show the observed rotation curves. The red line shows the total rotation velocity, the dash-dotted line the dark matter, the dashed line the disc, and the dotted line the gas. Some galaxies have bulge component which is represented by the magenta dash-dotted line.}\label{fig:3}
	\end{center}
\end{figure*}

\subsection{Bayesian analysis}
\label{Bayesian}
We implement the Bayesian analysis by using the affine-invariant MCMC ensemble sampler in $emcee$ 
\cite{Foreman2013}. The posterior probability of parameter space is given by
\begin{eqnarray}
\nonumber P(V_{200},C_{200},\Upsilon_{\rm d},\Upsilon_{\rm b},\delta_{\rm D},\delta_{\rm i}|\mathrm{SPARC})=\mathcal{L}(V_{200},C_{200},\\\Upsilon_{\rm d},\Upsilon_{\rm b},\delta_{\rm D},\delta_{\rm i}|\mathrm{SPARC})
P(V_{200},C_{200},\Upsilon_{\rm d},\Upsilon_{\rm b},\delta_{\rm D},\delta_{\rm i}),
\end{eqnarray}
where the likelihood is derived from the $\chi^2$ function, $\mathcal{L}\sim e^{-\chi^2/2}$ and
\begin{eqnarray}\label{eq:chi2}
\chi^2=\sum_{k=1}^N \left(\frac{V_{\rm tot}(R'_{\rm  k};V_{200},C_{200},\Upsilon_{\rm d},\Upsilon_{\rm b},\delta_{\rm D})-V'_{\rm obs,k}}{\delta V_{\rm obs,k}^{'}}\right)^2,
\end{eqnarray}
where $N$ is the number of data point for individual galaxy, the observed rotation curve and its uncertainty at the radius $R_{\rm k}$ has been changed to $V'_{\rm obs,k}$ and $\delta V'_{\rm obs,k}$, with a dimensionless factor $\delta_{\rm i}$. The total rotation velocity $V_{\rm tot}$ at the radius $R'_{\rm k}$ is predicted by the halo parameters $\{V_{200},C_{200}\}$ and the galactic parameters $\{\Upsilon_{\rm d},\Upsilon_{\rm b},\delta_{\rm D}\}$. The prior probability is the product of respective priors, 
\begin{equation}
P(V_{200},C_{200},\Upsilon_{\rm d},\Upsilon_{\rm b},\delta_{\rm D},\delta_{\rm i})=P(V_{200})P(C_{200})P(\Upsilon_{\rm d})P(\Upsilon_{\rm b})P(\delta_{\rm D})P(\delta_{\rm i}).
\end{equation}
Similarly to Li
et al. \cite{Li:2018rnd}, we impose the same priors on galactic parameters: Gaussian priors on $\delta_{\rm D}$ and $\delta_{\rm i}$ around 1 with standard deviations given by the observational relative errors; log-normal prior on $\Upsilon_*$ around their fiducial values $\Upsilon_{\rm d}=0.5$ and $\Upsilon_{\rm b}=0.7$ with a standard deviation of 0.1 dex suggested by the stellar population synthesis models. For halo parameters $\{V_{200},C_{200}\}$, a flat prior is used with $10<V_{200}<500~\mathrm{km\,s^{-1}},1<C_{200}<100$. These loose priors have also been used to fit the DC14 and NFW profile in Katz et al. \cite{2017MNRAS.466.1648K}. Moreover, we use flat priors because we want to compare our results with the scaling relations in D09, G09 and Kormendy $\&$ Freeman \cite{Kormendy2004,Kormendy2016} that they obtained by means of flat priors. Finally, the best fitting value is obtained by maximizing the posterior probability.

\section{Result}
\label{Result}


\subsection{Best fit to individual galaxy}
\label{Fit}

Based on the Bayesian method, we have fit 175 individual galaxy rotation curve in SPARC sample with the Burkert profile.
Fig. \ref{fig:3} shows the best fit to three representative galaxies,
with the meaning of the symbols described in the caption of the figure. 
We find that the Burkert profile can give a good fit to the galaxies studied.
%
%
The reduced $\chi^2$ for the galaxies in Fig. \ref{fig:3} is 0.867, 2.631, 0.988, respectively.	

Except {for} five galaxies, the galaxies have a reduced $\chi^2 <10$.
The best fitting values and the reduced $\chi^2$ for the full SPARC sample are listed in Table \ref{sample}.

In the rest of the paper, we will show the plots obtained using the Burkert profile, since we want to compare our results to that of D09, and G09.

\subsection{Correlations between halo and disc properties}
\label{Correl}

{In our paper, we studied the
	correlations between the product of the scale radius, $r_0$, and the central density, $\rho_0$, 
	with luminosity $L_{[3.6]}$.} 
We also showed the correlations between the dark matter acceleration $g_{\rm DM}(r_0)$ at scale radius
with luminosity $L_{[3.6]}$, which will be analyzed later.
D09 considered the correlation between $\rho_0 r_0$, and the galaxy magnitude, $M_B$. Here instead of $M_B$, we use $L_{[3.6]}$, since in the SPARC sample instead of $M_B$, the luminosity $L_{[3.6]}$ is used. 
Apart those correlations, we also studied, the correlations between $\rho_0 r_0$, {$g_{\rm DM}(r_0)$}, and other galaxies disc properties. In the present paper, we are mainly interested in studying 
the $\rho_0 r_0$ correlations. This because we want to compare the results with those of D09, and G09. 
As we already reported, D09, and G09 claimed a quasi-universality of $\rho_0 r_0$, where $r_0$ are the scale radius for {the} Burkert profile, and $\rho_0$ the central density. G09 
extended the result to baryons, claiming that the D09
result was valid for luminous matter surface density. In the following, we will show  
that fitting SPARC data by means of {the} Burkert profile like that used by D09, and G09, we do not find any quasi-universal relation, in agreement with several papers in literature \cite{Boyarsky,Napolitano2010,CardoneTortora2010,CardoneDelPopolo2012,2013MNRAS4291080D,Saburova2014,Li:2018rnd}.


\begin{table*}
	\begin{center}
		\begin{threeparttable}
			\caption[]{The linear regression between the dark halo surface density and the galactic disc properties. (1) -- the linear regression equation; (2) -- the Pearson correlation coefficient $R$; (3) -- the significance level\textsuperscript{*}.}\label{tab1}
			\begin{tabular}{lrc}
				\hline
				Equation&\multicolumn{1}{c}{$R$}&$n\sigma$\\
				(1)&\multicolumn{1}{c}{(2)}&\multicolumn{1}{c}{(3)}\\
				\hline
				$\log_{10} \rho_0r_0 = (0.13 \pm 0.02)\log_{10} L_{[3.6]}          + (0.95 \pm 0.23)$& 0.33& 4.45$\sigma$\\
				$\log_{10} \rho_0r_0 =-(0.07 \pm 0.01)T                            + (2.72 \pm 0.07)$&-0.42& 5.81$\sigma$\\
				$\log_{10} \rho_0r_0 = (0.90 \pm 0.12)\log_{10} V_{\rm{flat}}      + (0.42 \pm 0.25)$& 0.54& 6.81$\sigma$\\
				$\log_{10} \rho_0r_0 = (0.46 \pm 0.08)\log_{10} R_{\rm{eff}}       + (2.08 \pm 0.04)$& 0.15& 1.94$\sigma$\\
				$\log_{10} \rho_0r_0 = (0.25 \pm 0.04)\log_{10} \Sigma_{\rm{eff}}  + (1.73 \pm 0.09)$& 0.50& 7.07$\sigma$\\
				$\log_{10} \rho_0r_0 = (0.32 \pm 0.08)\log_{10} R_{\rm{disc}}      + (2.17 \pm 0.04)$& 0.14& 1.88$\sigma$\\
				$\log_{10} \rho_0r_0 = (0.36 \pm 0.04)\log_{10} \Sigma_{\rm{disc}} + (1.27 \pm 0.11)$& 0.51& 7.21$\sigma$\\
				$\log_{10} \rho_0r_0 = (0.32 \pm 0.08)\log_{10} R_{\rm{HI}}        + (1.90 \pm 0.10)$& 0.14& 1.84$\sigma$\\
				$\log_{10} \rho_0r_0 = (0.22 \pm 0.04)\log_{10} M_{\rm{HI}}        + (0.20 \pm 0.39)$& 0.20& 2.67$\sigma$\\
				\hline
			\end{tabular}
			\begin{tablenotes}
				\small
				\item \textsuperscript{*}$n=\sqrt{2}\erf^{-1}(1-p)$, where $p$-value is the probability that the real data is reproduced by an uncorrelated system.
			\end{tablenotes}
		\end{threeparttable}
	\end{center}
\end{table*}

All the figures in this paper shows the quantity $\rho_0 r_0$ on the left vertical axis, and $g_{\rm DM}(r_0)$ on the right vertical axis in order to reduce the number of figures. This can be done since $\rho_0 r_0$, and $g_{\rm DM}(r_0)$ are {proportional} as shown in the next sections.

\begin{figure}
	\begin{center}
		\includegraphics[width=\linewidth]{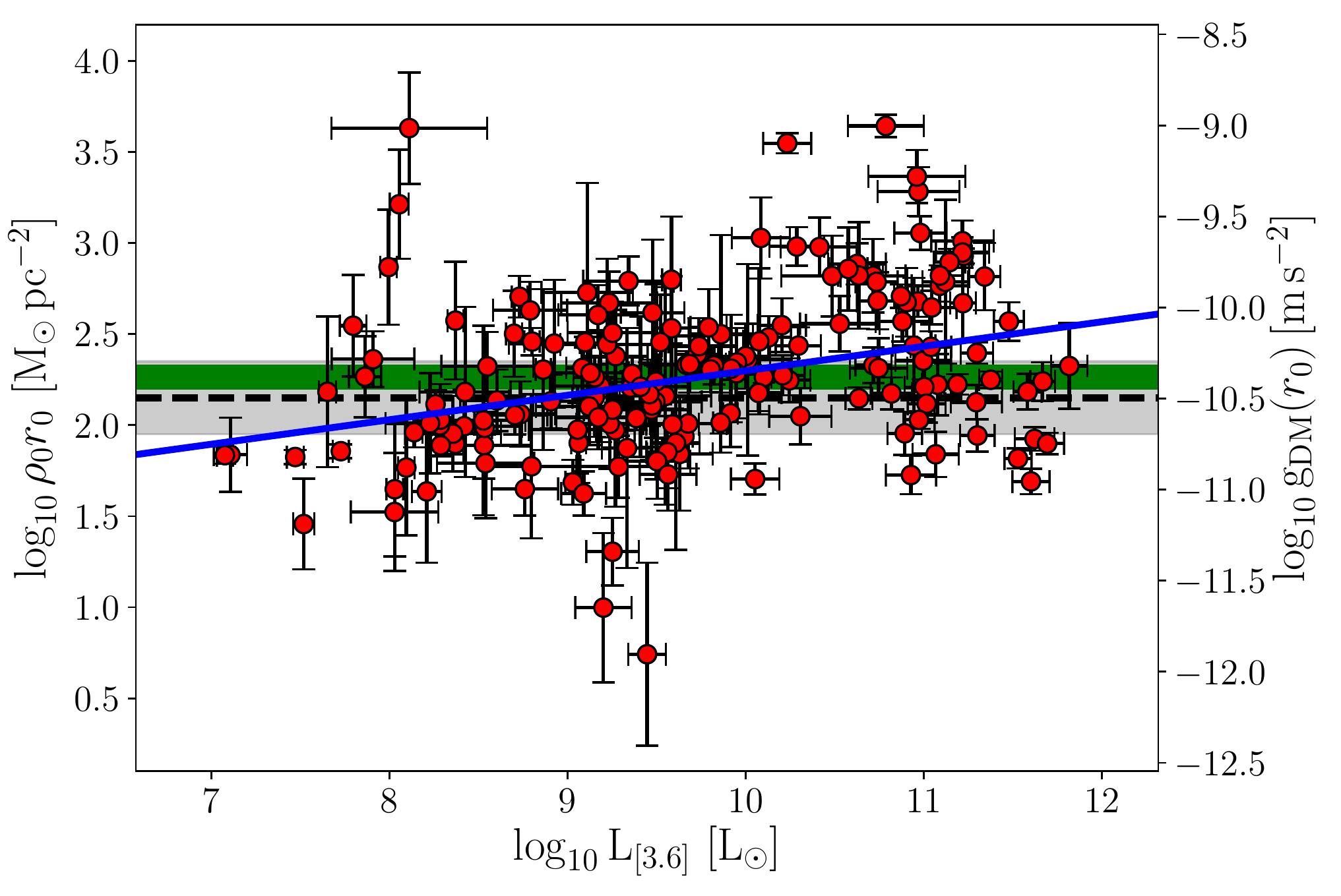}
		\caption{
			Scaling relations between the $\rho_0r_0$ (left vertical axis), $g_{\rm DM}(r_0)$ (right vertical axis), and $L_{[3.6]}$ for {the} Burkert profile when imposing flat prior on the halo parameters. The dashed black line, and the gray shaded region, represent the value 
			$\log_{10}{\rho_0 r_0}= 2.15 \pm 0.2$ obtained by D09.
			The blue line denotes the linear regression in log-space.
			The green shaded region, represents the range $0.3a_0-0.4a_0$ for the maximum halo acceleration predicted by Milgrom $\&$ Sanders \cite{Milgrom2005}.	
			%
		}
		\label{fig:R0}
	\end{center}
\end{figure}

In Fig. \ref{fig:R0}, we show the scaling relation between $\rho_0r_0$ and $L_{[3.6]}$, when imposing flat priors {on the halo parameters}.
Errors on  $\rho_0 r_0$, were obtained by error propagation {based on the uncertainties in the fitting parameters}, while the uncertainty on $L_{[3.6]}$ are obtained from SPARC dataset, and is given by the quadratic sum of errors on distances and flux.
The blue line in Fig \ref{fig:R0} denotes the linear regression in log-space.
%
%
The horizontal dashed line and the gray shaded region represent the predictions of D09 concerning the surface density, namely 
$\log_{10}{\rho_0 r_0}=2.15 \pm 0.2$ in units of ${\rm M_{\odot}\,pc^{-2}}$.
{The correlation strength is evaluated through the Pearson correlation coefficient 
$R$ and the $p$-value. The second one is the probability that the real data is reproduced by an uncorrelated system. Since the $p$-value is small, we convert it to the number of $\sigma$	from zero, $n=\sqrt{2}\erf^{-1}(1-p)$, where $\erf^{-1}$ is the inverse error function. Then the significance level for correlation is $n\sigma$. 
In this scaling relation, $R=0.33$, and we find that the significance level for correlation is $4.45\sigma$, indicating a medium correlation between the two variables. This result is shown in Table \ref{tab1}.}
The smaller value of the Pearson coefficient $R$ in our case with respect to Li et al.
\cite{Li:2018rnd} is due to the flat prior on halo parameters we used. It seems that Gaussian priors used in Li et al. \cite{Li:2018rnd} impose a strong constraints on the halo parameters.
We fit the data by using the linear regression and we find a linear relation in log-space,
\begin{equation}
\log_{10} \rho_0r_0 = (0.13 \pm 0.02)\log_{10} L_{[3.6]}          + (0.95 \pm 0.23).
\end{equation}
The quoted result is in contradiction to that 
of D09, G09, and Kormendy $\&$ Freeman \cite{Kormendy2004,Kormendy2016}, which does not show any correlation (flat line). The dark halo profile we used for our fit is the same of that of D09, 
and G09,
while Kormendy $\&$ Freeman \cite{Kormendy2016} used a non-singular isothermal sphere. In the present paper, we are interested in a comparison with D09, so we are not going to consider the changes due to non-singular isothermal sphere, that however should be small since the Burkert profile, excluding the outer regions, has a similar behavior to the non-singular isothermal sphere. 

%
%
%
%

\begin{figure*}
	\begin{center}
		\includegraphics[width=0.44\textwidth]{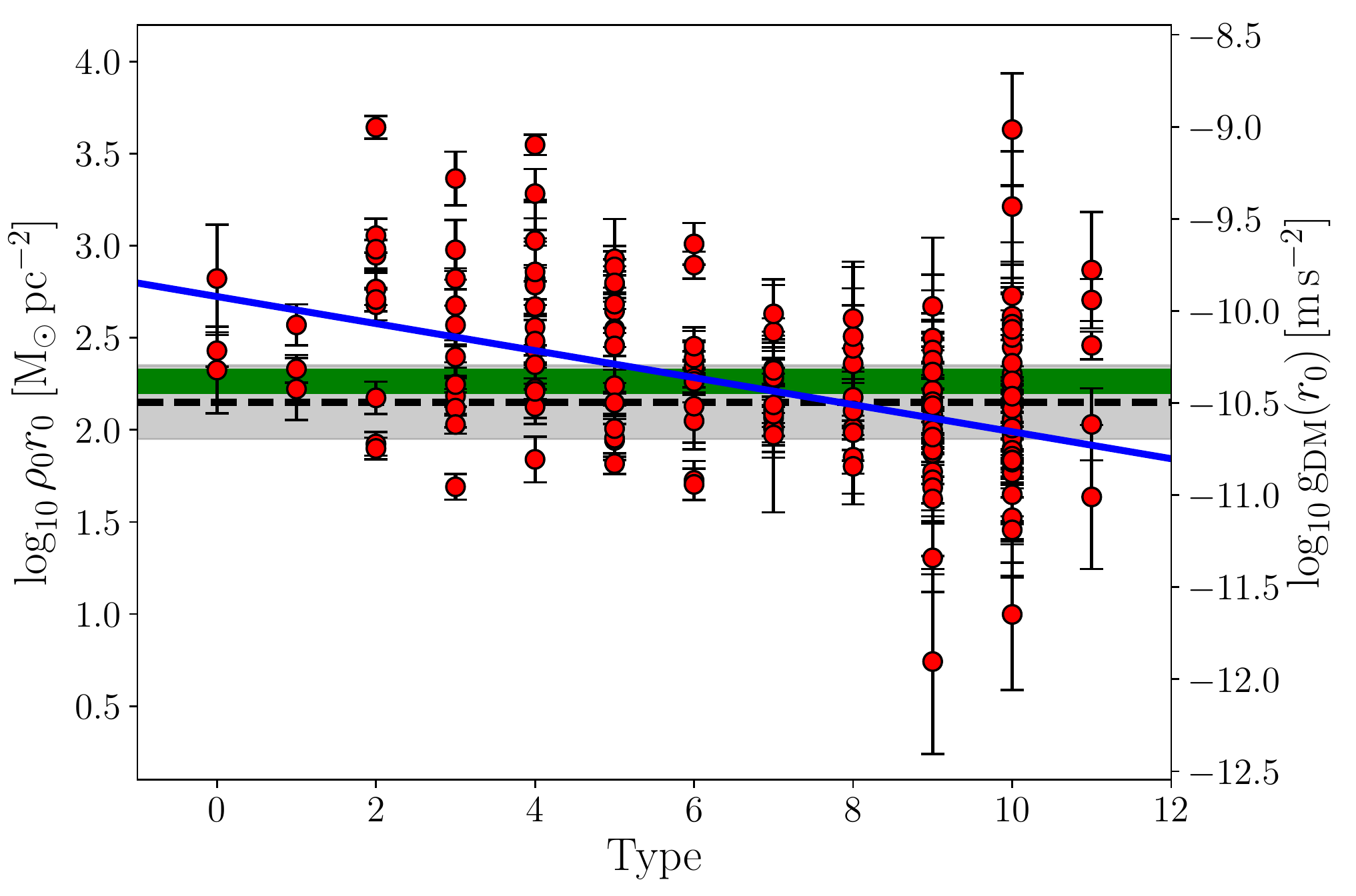}
		\includegraphics[width=0.44\textwidth]{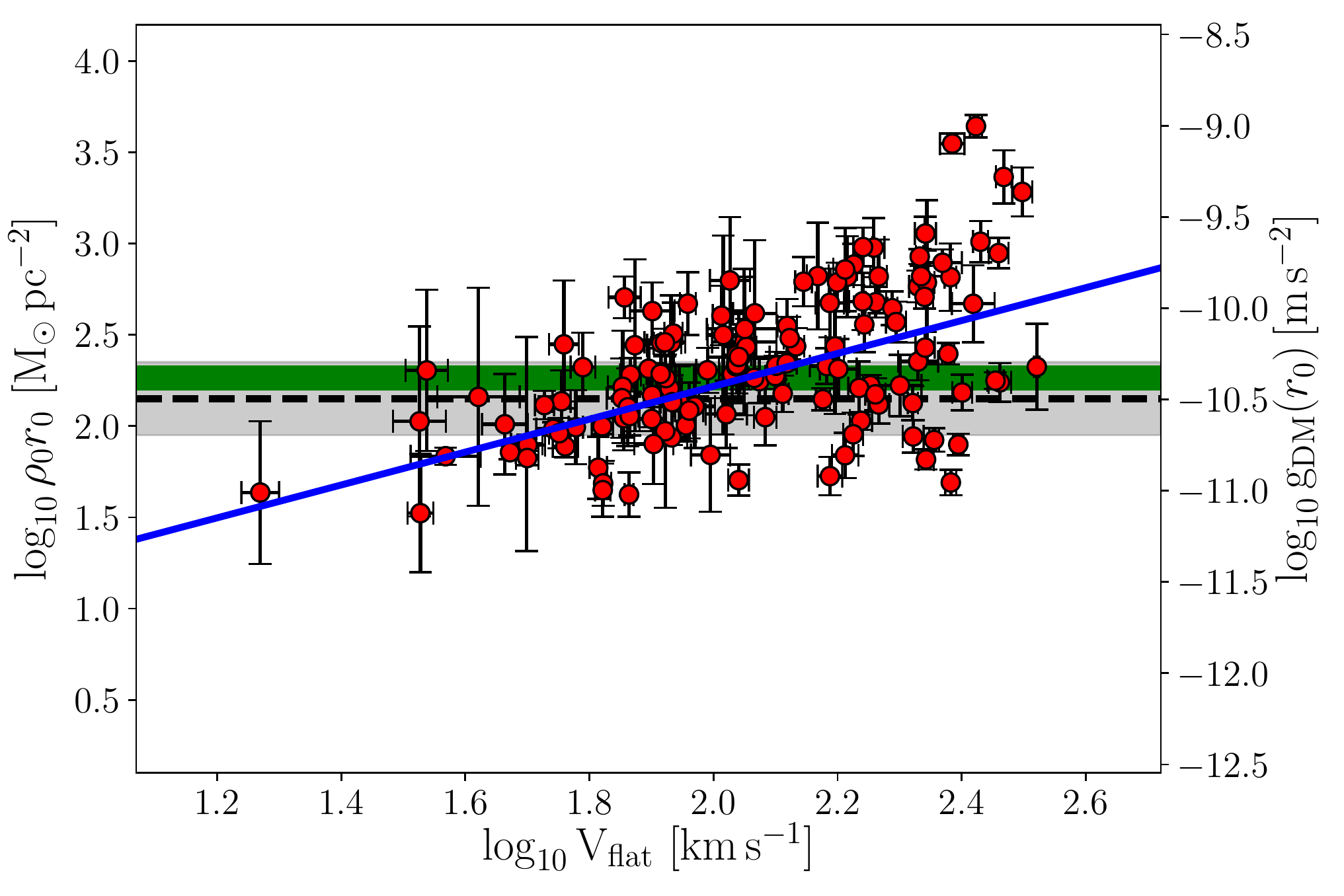}
		\includegraphics[width=0.44\textwidth]{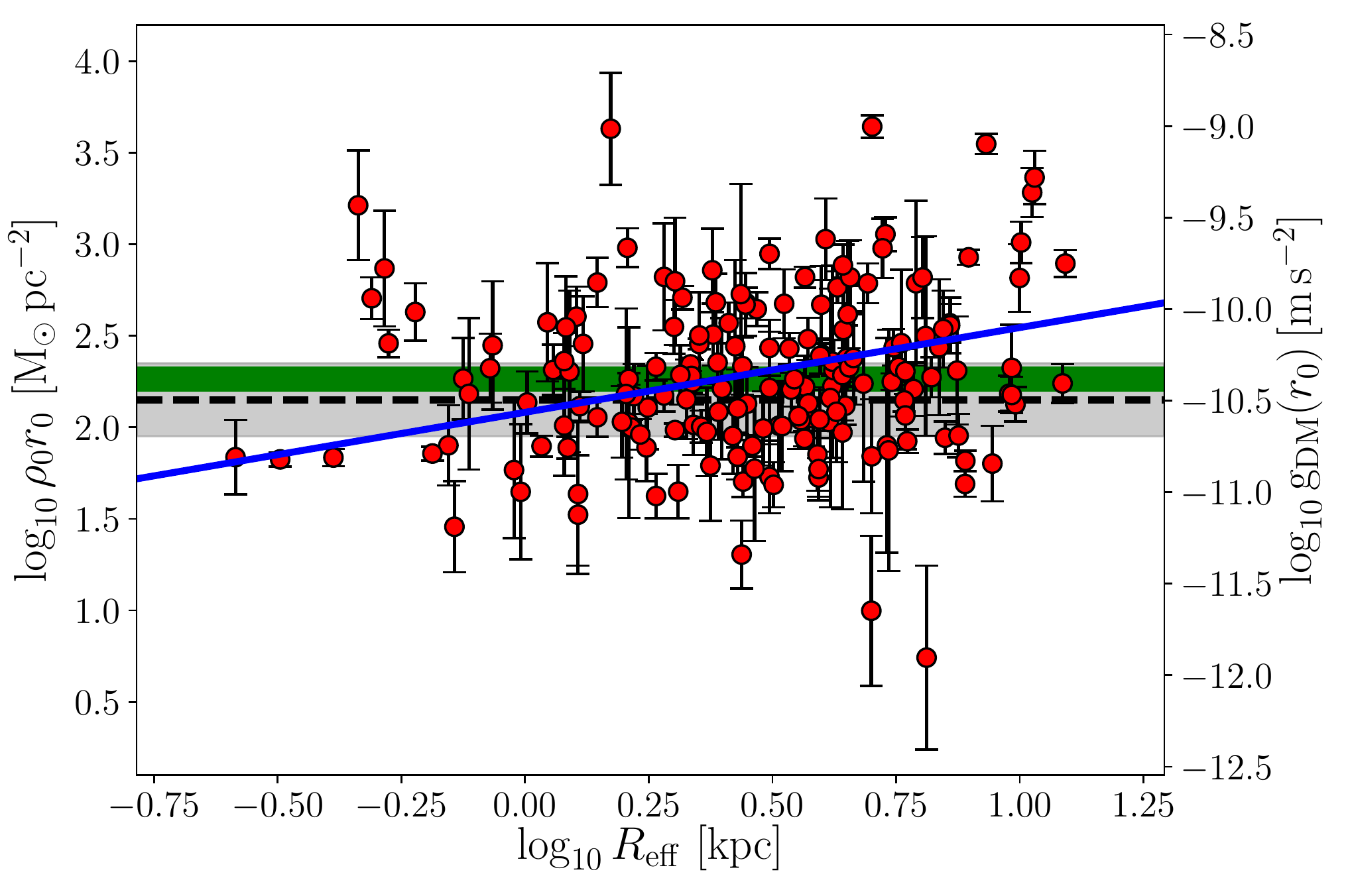}	
		\includegraphics[width=0.44\textwidth]{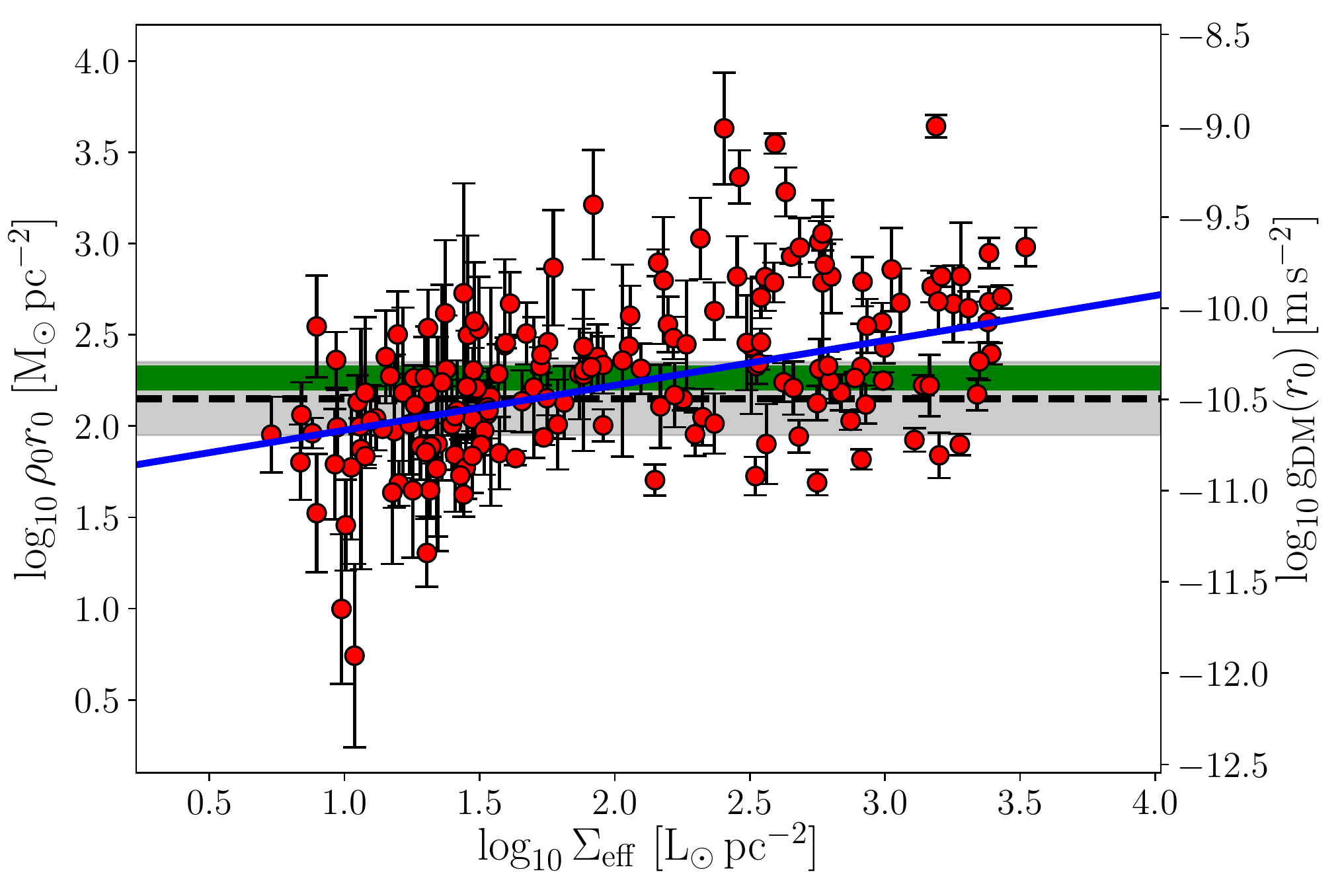}
		\includegraphics[width=0.44\textwidth]{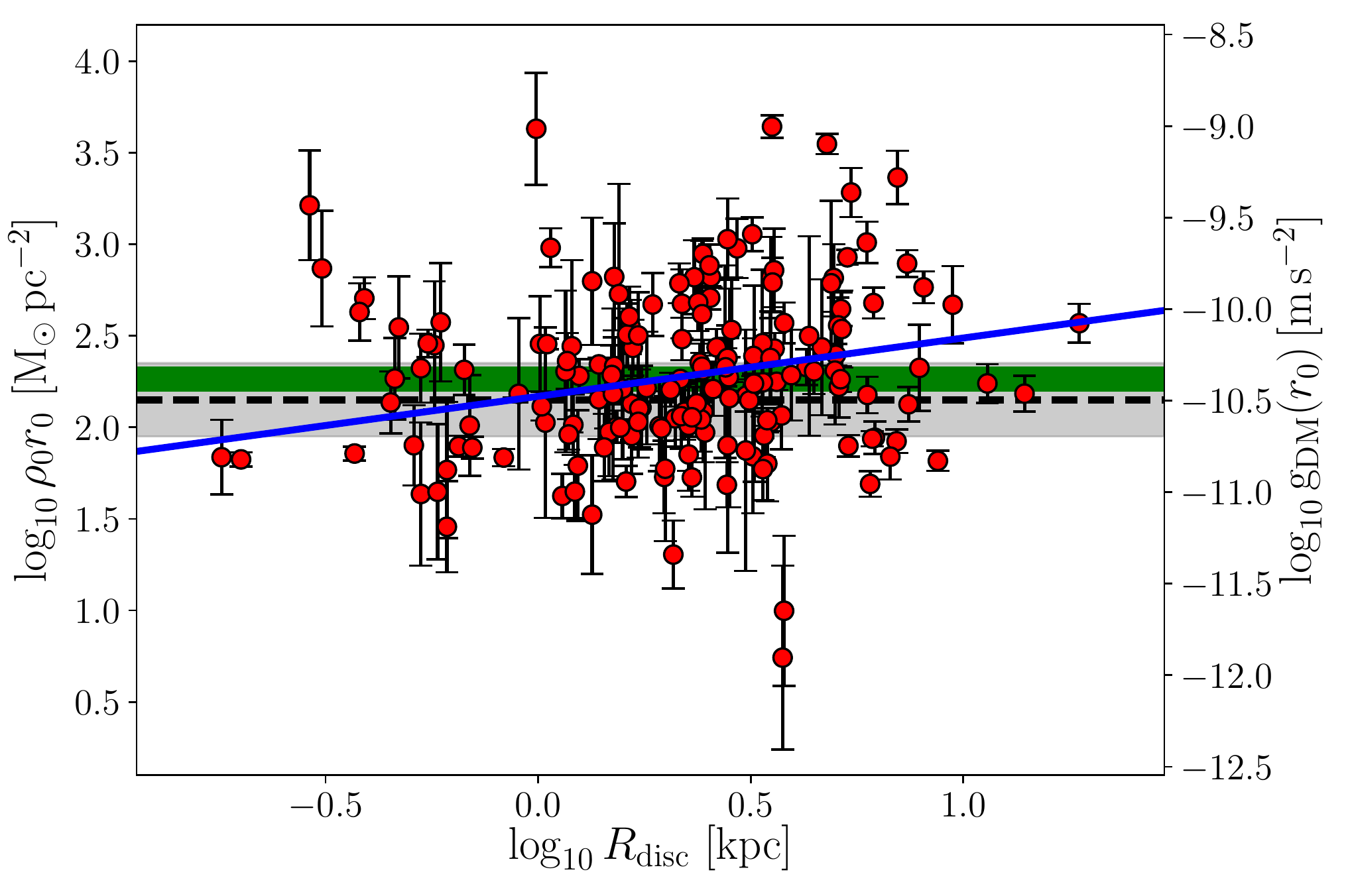}
		\includegraphics[width=0.44\textwidth]{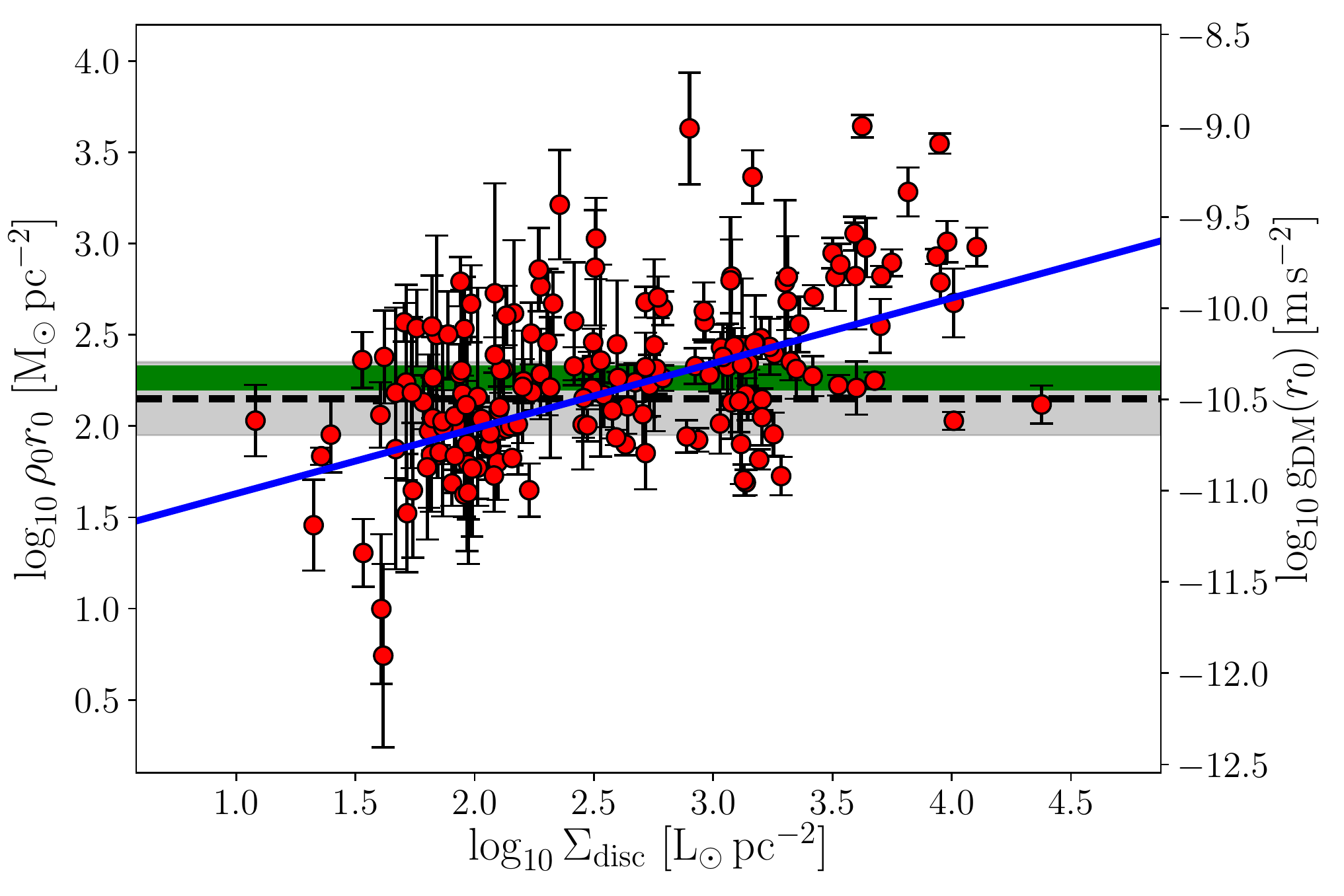}
		\includegraphics[width=0.44\textwidth]{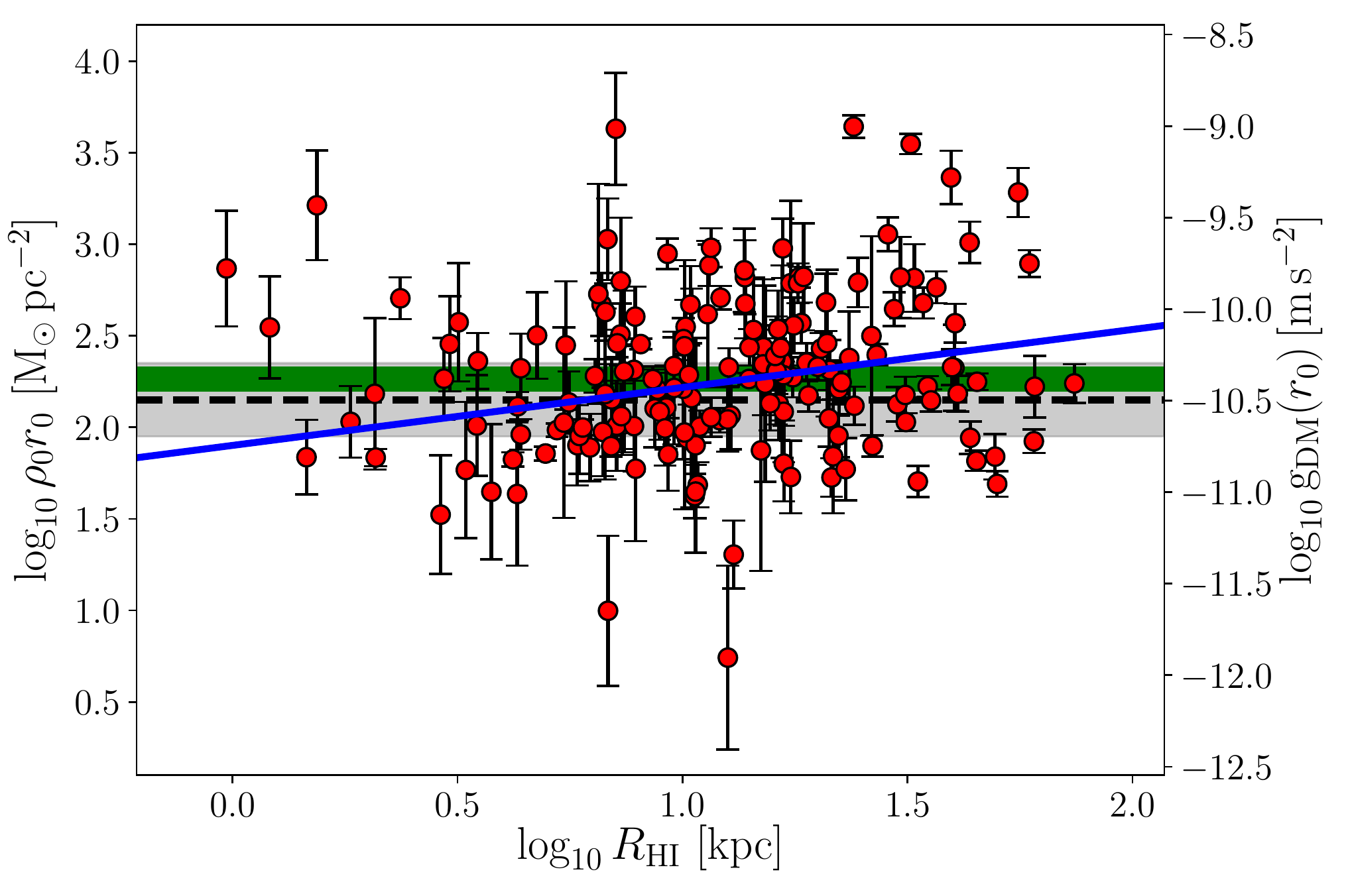}		
		\includegraphics[width=0.44\textwidth]{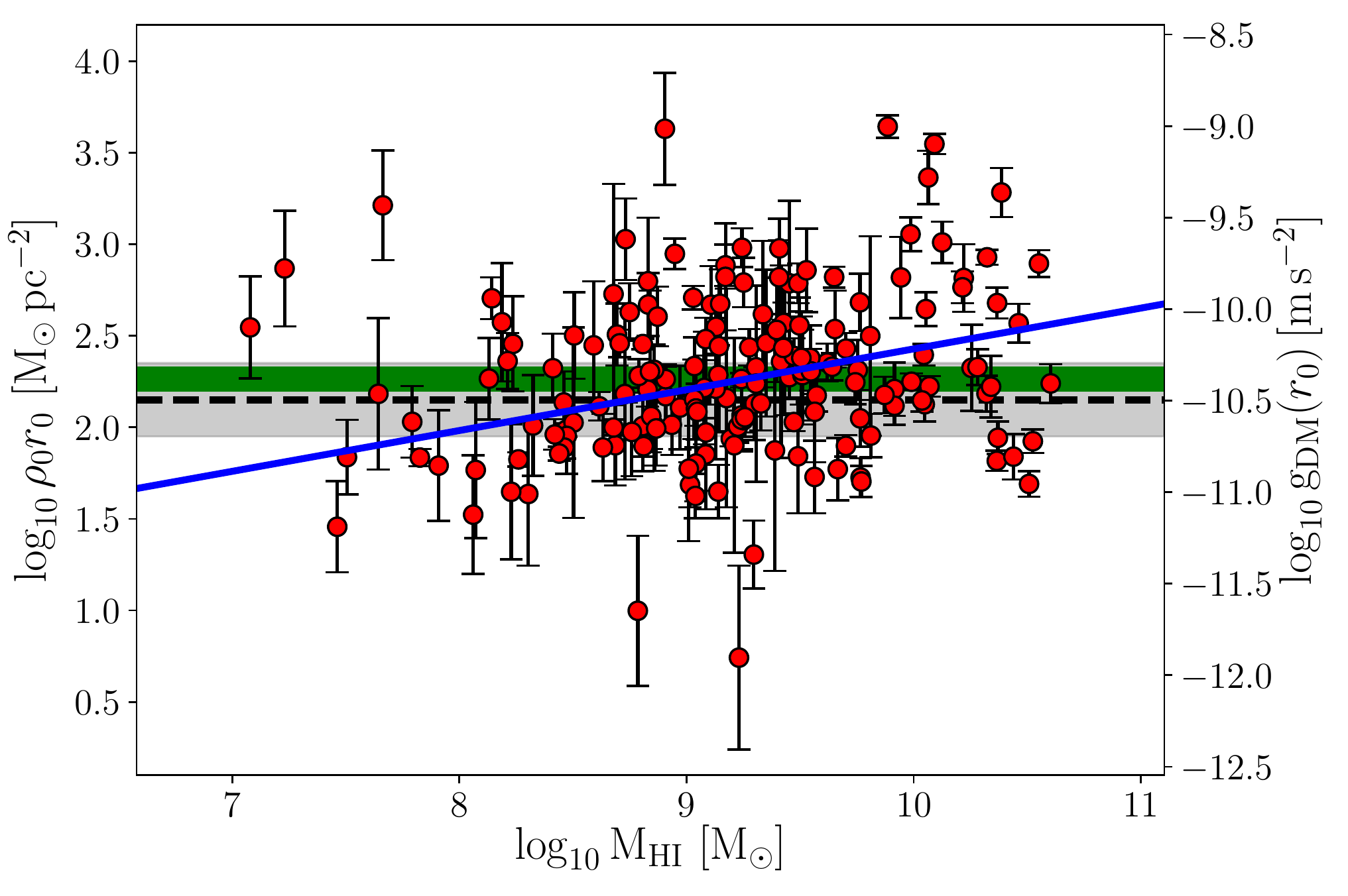}
		\caption{
			Scaling relations between the $\rho_0r_0$ (left vertical axis), $g_{\rm DM}(r_0)$ (right vertical axis)
			vs galactic properties. The blue line denotes the linear regression in log-space (except the Hubble type). The dashed black line, and the gray shaded region, represent the value 
			$\log_{10}{\rho_0 r_0}= 2.15 \pm 0.2$ obtained by D09.
			The green shaded region, represents the range $0.3a_0-0.4a_0$ for the maximum halo acceleration predicted by Milgrom $\&$ Sanders \cite{Milgrom2005}.	
			Circles with error bars correspond to the 
			data obtained by means of the SPARC sample. The panels show the correlations: 1. $\rho_0r_0$($g_{\rm DM}$) -- numerical Hubble type; 2. $\rho_0r_0$($g_{\rm DM}$) -- rotation velocity along the flat part; 3. $\rho_0r_0$($g_{\rm DM}$) -- effective radius; 4. $\rho_0r_0$($g_{\rm DM}$) -- effective surface brightness; 5. $\rho_0r_0$($g_{\rm DM}$) -- scale length of the stellar disc; 6. $\rho_0r_0$($g_{\rm DM}$) -- central surface brightness of the stellar disc; 7. $\rho_0r_0$($g_{\rm DM}$) -- HI radius; 8. $\rho_0r_0$($g_{\rm DM}$) -- total HI mass.
			%
		}
		\label{fig:T}
	\end{center}
\end{figure*}

As we already discussed, we used {the} Burkert profile in order our analysis is similar to that of D09. 
A comparison of our result, Fig. \ref{fig:R0}, with their result shows they do not agree. While D09 do not find any correlation between $\rho_0 r_0$ and magnitude (luminosity), in our case there is a correlation similarly to previously cited papers \cite{Boyarsky,Napolitano2010,CardoneTortora2010,CardoneDelPopolo2012,2013MNRAS4291080D,Saburova2014,Li:2018rnd}.

So summarizing, following the method used by D09, we found a result contradicting theirs. 
Apart the correlation, Fig. \ref{fig:R0} shows that our data are not contained in the D09 range, showing again that D09 result is not in agreement with ours: there is no hint of a quasi-universal behavior of $\rho_0 r_0$, and this conclusion is again in agreement with several papers \cite{Boyarsky,Napolitano2010,CardoneTortora2010,CardoneDelPopolo2012,2013MNRAS4291080D,Saburova2014}. 

We want to add that D09 analysis has some issues. As noticed by Li et al. \cite{Li:2018rnd}, in D09 stellar contributions is taken into account using several methods {and adopting spectro-photometric galaxy models. Thus, the contributions of each component strongly depend on the efficacy of the modeling.} Moreover, for several galaxies, D09 did not fit the rotation curve using {the} Burkert profile but relied on values already presented in literature, and based on different dark halo profiles. We improved the previous issues by fitting each SPARC galaxy rotation curve to obtain the Burkert parameters, using 
the MCMC method, also to infer a realistic estimate of the errors on the quantities of interest. 

%
%


The analysis in Spano et al. \cite{Spano2008} and Kormendy $\&$ Freeman \cite{Kormendy2016} also shows some inappropriate choices.
As noticed by Li et al. \cite{Li:2018rnd}, and  previously reported by D09, the use of the maximum disc analysis by Kormendy $\&$ Freeman \cite{Kormendy2004,Kormendy2016} produces drawbacks, as pushing $\Upsilon_{\rm disc}$ to  unreasonably high values in the case of low mass galaxies. Spano et al. \cite{Spano2008} assumed constant 
$\Upsilon_{\star}$, while in the optical band a strong variation of $\Upsilon_{\star}$ is expected \cite{McGaugh2014}. 
%
%

\subsection{Other correlations}
\label{Other}

In order to show that the correlation we find is a genuine one, we looked for other correlations between $\rho_0 r_0$ and the galaxy disc properties as tabulated in the SPARC data set \cite{Lelli:2016zqa}. The result of the analysis is plotted in Fig. \ref{fig:T}. 
{The first correlation we found is with galaxy type, characterized by $R=-0.42$, {having a significance level of $5.81\sigma$}, 
	and}
\begin{eqnarray}
\log_{10} \rho_0r_0 =-(0.07 \pm 0.01)T                       + (2.72 \pm 0.07).
\end{eqnarray}  
It shows that earlier galaxies have a larger surface density.
%
%

The second one is with the rotation velocity along the flat part (only 135 galaxies have the flat part), $V_{\rm flat}$, {characterized by $R=0.54$, indicating a strong correlation with a significance level of $6.81\sigma$,} and
\begin{eqnarray}
\log_{10} \rho_0r_0 = (0.90 \pm 0.12)\log_{10} V_{\rm{flat}}      + (0.42 \pm 0.25).
\end{eqnarray}

We also found that there exist strong correlations with the baryonic surface brightness, which can 
be converted into surface density via the mass-to-light ratio. 
Also the effective surface brightness, $\Sigma_{\rm eff}$, and the central surface brightness of the stellar disc, $\Sigma_{\rm disc}$ can be expressed in terms of the dark halo surface density, strongly correlated to $\rho_0 r_0$.
%
%
There still exist positive correlations with the effective radius, $R_{\rm eff}$, and the scale length of the stellar disc, $R_{\rm disc}$, but the correlations are weak. We also find other two correlations with the total mass of atomic hydrogen (HI), $M_{\rm HI}$, and with the radius where the HI surface density reaches 1 ${\rm M_{\odot}\,pc^{-2}}$. These results are summarized in Table \ref{tab1}.


{In all panels, the horizontal dashed line and the gray shaded band represent the predictions of D09 concerning the surface density, namely $\log_{10}{\rho_0 r_0}=2.15 \pm 0.2 {\rm \,M_{\odot}\,pc^{-2}}$.}
%
%

Summarizing, in our analysis we followed closely D09 analysis, and differently from them we found a 
correlation between $\rho_0 r_0$ and luminosity, in agreement with several previous studies \cite{Boyarsky,Napolitano2010,CardoneTortora2010,CardoneDelPopolo2012,2013MNRAS4291080D,Saburova2014,Li:2018rnd}. Moreover, our data do not fulfill the D09 claim that $\log_{10}{\rho_0 r_0}=2.15 \pm 0.2 {\rm \,M_{\odot}\,pc^{-2}}$.

The main point of the previous plots is that the quoted correlations 
leave small room to the idea that $\rho_0 r_0$ is constant. 

\begin{figure}
	\begin{center}
		\includegraphics[width=\linewidth]{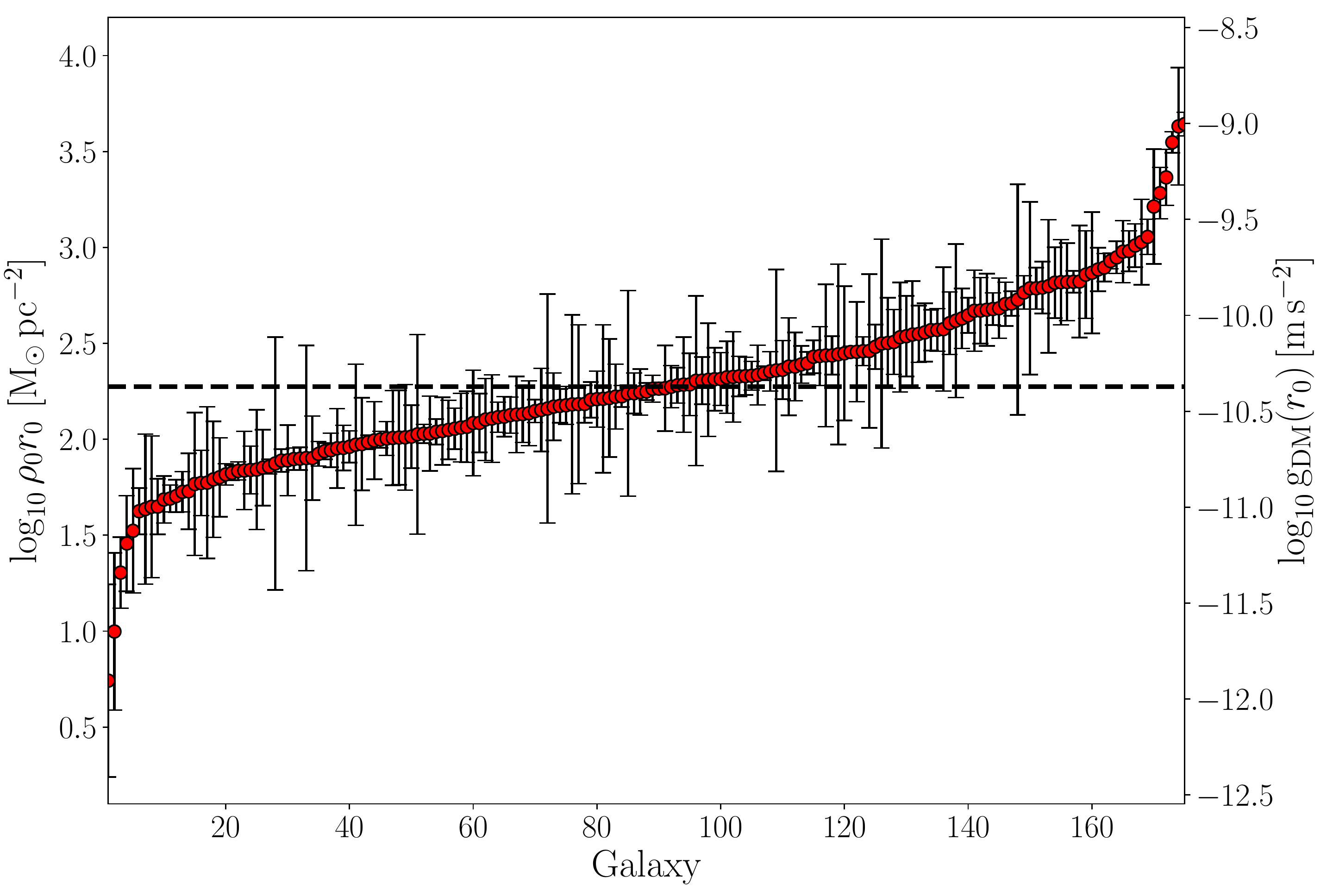}
		\caption{The derived posterior probability distributions of ${\rho_0 r_0}$ (left vertical axis) and $g_{\rm DM}(r_0)$ (right vertical axis) for 175 SPARC galaxies. The red circles show the maximum of the posterior {in ascending order}. The error bars show its uncertainties. The global best-fit value of $\rho_0 r_0$, $g_{\rm DM}(r_0)$ is shown by the dashed line. 
		}		
		\label{poster}
	\end{center}
\end{figure}

In order to have a more quantitative idea, following Rodrigues et al. \cite{Rodrigues:2018duc},
we calculated the confidence level to reject the hypothesis of a constant value of $\rho_0 r_0$. Fig. \ref{poster} show the derived posterior probability distributions of ${\rho_0 r_0}$ for 175 SPARC galaxies. Each red circle shows the maximum of the posterior, and the dashed line is the global best-fit of ${\rho_0 r_0}$. We find that the global best-{fit} value is  $\log_{10}{\rho_0 r_0}=2.27$, i.e., $\rho_0 r_0=187.94\,\rm M_{\odot}\,pc^{-2}$. Most of galaxies are quite incompatible with the global best-fit. The null hypothesis (constancy of the $\rho_0 r_0)$ is rejected at $>10 \sigma$.
After excluding low-quality galaxies following the quality criteria in Rodrigues et al. \cite{Rodrigues:2018duc}, namely using the same 100 galaxies, we find that the null hypothesis is still rejected at 
$>10 \sigma$.


\section{The surface density, DM Newtonian acceleration and MOND}
\label{MOND}

As several times reported, D09 found that the product $\Sigma^0_{\rm Donato} = \rho_0 r_0$, dubbed surface density, is constant, independent on $M_B$ magnitude (i.e., luminosity), in a very large magnitude range: $-8 \geq M_B \geq -22$, having the value,

\begin{eqnarray}
\log_{10}\frac{\Sigma^0_{\rm Donato}}{\rm M_{\odot}\,pc^{-2}}=2.15 \pm 0.2.
\label{eq:don}
\end{eqnarray}
From Eq. \ref{eq:don}, one gets the gravitational acceleration coming from the dark matter component at the scale radius
\begin{eqnarray}
g_{\rm dark}(r_0)= G \pi \Sigma_{\rm dark}(r_0)=3.2^{+1.8}_{-1.2}\, 10^{-9}\, \rm cm\,s^{-2},
\end{eqnarray}
where $\Sigma_{\rm dark}(r_0)=0.51\Sigma^0_{\rm Donato}$ is the dark matter mean surface density within the scale radius $r_0$. This acceleration assumes always the same value.

G09 extended this result to the luminous component in galaxies,
\begin{eqnarray}
g_{\rm bary}(r_0) = G \pi \Sigma_{\rm bary}(r_0),
\end{eqnarray}
{where $\Sigma_{\rm bary}(r_0)$ is the baryonic mean surface density within $r_0$} and they found $g_{\rm bary}(r_0)=5.7^{+3.8}_{-2.8}\, 10^{-10}\, \rm cm\,s^{-2}$.

This implies that a. the gravitational acceleration coming from the DM component, and the luminous galaxies component at the scale radius assumes always the same value; b. in a scale radius ($r_0$) the luminous-to-dark matter ratio,
is constant; c. the central baryonic surface density correlates with the core radius. 
The quoted claims were interpreted as a ``correlation" between the {enclosed surface densities of luminous and dark matter in galaxies}. Another interpretation that was given is that the DM halo core radius is the radius beyond which $g_{\rm bary}(r) <6 \times 10^{-10}\, \rm cm\,s^{-2}$, relating the result to the so called mass 
discrepancy-acceleration relation \cite{1983ApJ270365M}. 
From the ``universal and maximum" acceleration $g_{\rm dark}(r_0)$, 
the mass discrepancy-acceleration relation predicts, by definition, the existence of a universal gravity coming from baryons at $r_0$, and a universal surface density in $r_0$. 

In other words, there is a strict correlation between the constancy (universality) of the surface density and a universal acceleration, that is nothing else than the well known universal acceleration in MOND, $a_0$.

Milgrom \cite{Milgrom2009} prompted by the previous results showed that MOND predicts, for all object having a mean acceleration at or above $a_0$, a quasi-universal central surface density of galaxy DM haloes, which can be written as 

\begin{eqnarray}
\Sigma^0_{\rm MOND}=\gamma \Sigma_M=\gamma \frac{a_0}{2 \pi G},
\label{eq:Siao}
\end{eqnarray}
where $0.7<\gamma < 2$, for the limiting form of the interpolating function used in \cite{Milgrom2009},
and  

\begin{equation}
\Sigma_{M}=138 \frac{a_0}{\rm 1.2 \times 10^{-8}\,cm\, s^{-2}}{\rm M_{\odot}\,pc^{-2}},
\end{equation}
that for a canonical value of $a_0$ gives $\Sigma_M=138\, {\rm M_{\odot}\,pc^{-2}}$, or 
$\log_{10}{\frac{\Sigma_M}{\rm M_{\odot}\,pc^{-2}}}=2.14$,
%
%
in agreement with Eq. \eqref{eq:don}. In the case of low surface density systems, having accelerations smaller than $a_0$, $\Sigma^0_{\rm MOND} \simeq 2.4 (\Sigma^0_b \Sigma_M)^{1/2} \simeq (6/\pi+1) \Sigma_{M} X_0 \simeq 0.6 \Sigma_{ M}$, 
%
%
being $X_0 \simeq 0.2$, and $\Sigma^0_b$ the central baryon surface density.
The previous result was generalized by \cite{Milgrom2016}, in $\Sigma^0_{\rm MOND}= \Sigma_M S(\Sigma^0_b/\Sigma_M)$, where the function $S(y)=\int^y_0 \nu(y') dy'$, and $\nu$ is the interpolating function. 
%
%
In summary, Milgrom \cite{Milgrom2009} confirms D09 result, adding that the quasi-universal value is not shared by objects with low surface densities, and that values lower than $\Sigma^0_{\rm Donato}$ are allowed in low surface density systems. To be conservative, D09 result together with Milgrom \cite{Milgrom2009} can be written as

\begin{eqnarray}
1.9<\log_{10} \Sigma^0_{\rm MOND} <2.4,
\label{eq:lim} 
\end{eqnarray}
which almost coincide with the range $1.95<\log_{10} \Sigma^0_{\rm Donato} <2.35$ \citep{Donato}.
Moreover, Eq. \eqref{eq:Siao} with Eq. \eqref{eq:lim} impose a limit on $a_0$, showing that there is a strict correlation between the surface density properties and MOND.

A spontaneous question, at this point is: are our data on $\Sigma^0_{\rm Donato} =\rho_0 r_0$ compatible with MOND predictions?

As shown in Fig. \ref{fig:R0}, the relation between $\Sigma^0_{\rm Donato} =\rho_0 r_0$ and the luminosity is not flat as predicted in D09, 
but show a precise correlation already discussed. The plot also shows that the distribution of the points in Fig. \ref{fig:R0} is largely outside the prediction of D09.
Fig. \ref{fig:T}, shows a similar behavior.
The relations between $\rho_0 r_0$ and the quantities in the {x-axis}, are never flat, and the points are largely distributed out from the D09 boundaries. 
Moreover, the condition Eq. \eqref{eq:lim}, is violated by our data, implying that Milgrom \cite{Milgrom2009} prediction obtained by means of MOND is violated by data, and this is a big problem for MOND. 

%

We can do another check using another MOND prediction given by Milgrom $\&$ Sanders \cite{Milgrom2005} concerning the existence of a maximum halo acceleration. This results comes out from the relation
\begin{eqnarray}
g_{\rm h}(g)=g-g_{\rm N}=g-g \mu(g/a_0)
\end{eqnarray}
where $g$ is the true MOND acceleration, $g_{\rm h}$ the halo acceleration, $g_{\rm N}$ the Newtonian acceleration, and $\mu$ is MOND interpolation function. {$g_{\rm h}$ cannot exceed $a_{\rm max}= \eta a_0$, with $\eta \simeq 0.3-0.4$ according to the standard interpolation function, $\mu(x)=\frac{x}{\sqrt{1+x^2}}$}.


We calculated,
from the plot of the correlation between the surface density and luminosity, Fig. \ref{fig:R0} (left axis \footnote{We write left axis to refer to $\rho_0 r_0$, and right axis to refer to $g_{\rm DM}(r_0)$.}),
the correlation between the acceleration and the luminosity, that we plotted in Fig. \ref{fig:R0} (right axis).
{From Fig. \ref{fig:R0} (right axis),
	it can be seen that our data do not confirm MOND prediction.}
In Fig. \ref{fig:T} (right axis),
we  also show the DM Newtonian acceleration vs galactic properties, showing the halo maximum accelerations predicted by MOND. 
{Following Milgrom $\&$ Sanders \cite{Milgrom2005}, we plot the band  $0.3a_0\sim0.4a_0$ (green area), indicating the range in which the maximum halo acceleration can be.
	Some of the galaxies have values of $g_{\rm DM}(r_0)$ that are not only higher than $0.4 a_0$
	but even $a_0$.}

%

The previous conclusion is strengthened by Fig. \ref{fig:T} (right axis)
showing that $g_{\rm DM}(r_0)$ is correlated with several quantities of the galactic disc, and again the plots show that the data {violates} the MOND \cite{Milgrom2005} predictions.

{Moreover, we follow the method introduced in Rodrigues	et al. \cite{Rodrigues:2018duc} to calculate the confidence level to reject a constant value of $g_{\rm DM}(r_0)$.  Fig. \ref{poster} (right axis)
	show the derived posterior probability distributions of $g_{\rm DM}(r_0)$ for 175 SPARC galaxies. We find that the global {best-fit} value is  $\log_{10}{g_{\rm DM}(r_0)}=-10.37$, i.e., $g_{\rm DM}(r_0)=0.42\times10^{-10}\,\rm m\,s^{-2}$. Note that even if the global {best-fit} of $g_{\rm DM}(r_0)$ is $0.35 a_0$, in the MOND prediction range  $0.3a_0\sim0.4a_0$, for most of galaxies, $g_{\rm DM}(r_0)$ is incompatible with the global {best-fit}. 
	The null hypothesis (constancy of the $g_{\rm DM}(r_0)$) is rejected at $>10 \sigma$.
	Even if we exclude those galaxies with low-quality following the quality criteria in Rodrigues	et al. \cite{Rodrigues:2018duc} and use the same 100 galaxies, we find that the null hypothesis is still rejected at  
$>10 \sigma.$
	
	%
	%

	\section{Discussion and Conclusions}
	\label{Conclusions}
	
	In this paper, using SPARC sample, we verified the D09 claim of the existence of a universal surface density of dark matter haloes. We calculated the $\rho_0 r_0$ {for the Burkert profile, by using the MCMC method.} We looked for correlations between the quoted quantity and luminosity, as done by D09, and also we verified if our $\rho_0 r_0$ data satisfied D09 result, namely $\log_{10}{\rho_0 r_0}=2.15 \pm 0.2 {\rm \,M_{\odot}\,pc^{-2}}$. We repeated the calculation, looking for correlation with a series of other disc properties (Hubble type, rotation velocity along the flat part, effective radius, effective surface brightness, scale length of the stellar disc, central surface brightness of the stellar disc, total HI mass and HI radius), finding similar results to that of the 
	$\rho_0r_0-L_{[3.6]}$ correlation. 
	The calculations, performed through Bayesian statistics, showed that contrarily to D09 conclusions, that the surface density is not a universal quantity.  
As shown by Milgrom \cite{Milgrom2009}, MOND has a strong prediction for the surface density. We verified if our data {is consistent with} that prediction, but we found the opposite result. To strengthen the previous result, we used another of the predictions of MOND \cite{Milgrom2005}, related to the existence of a maximum value for halo 
	Newtonian acceleration, $g_{\rm DM}(r_0)$, predicted to be in the range $0.3a_0\sim0.4a_0$ {for the standard interpolation function.}
	Also in this case, MOND predictions are in contradiction with data. The dark matter Newtonian acceleration correlates with all the previously presented galactic properties, and our calculated $g_{\rm DM}(r_0)$ is outside the boundary predicted by MOND. We also {calculated} the confidence level to reject a constancy of $g_{\rm DM}(r_0)$ and we find that the null hypothesis is rejected at a very high confidence level.
	
{The present paper is related to Rodrigues' \cite{Rodrigues:2018duc}. Like us, they use the SPARC database to show that the probability to have a fundamental acceleration (e.g., the typical MOND acceleration, $a_0$), is practically 0, and that $a_0$ must be of emergent nature. The analysis in the present paper, leads us to a similar conclusion. In this paper, we were not interested in studying what is the origin of the emergent nature of $a_0$. This was studied by a recent paper (Grudic et al. 2019,
arXiv:1910.06345),
which in agreement with our paper and \cite{Rodrigues:2018duc}, concludes that the origin of the quoted acceleration is emergent rather than fundamental. Moreover, according to their study, the acceleration comes naturally from stellar feedback. 
}	
	
	We may conclude that our results show the absence of a universal surface density, and the absence of maximum acceleration in haloes. Since these quantities are related to MOND, which has precise predictions for them, rejected by a comparison with data, this imply that MOND shows big problems at small scales, the scales at which should give its best.

\section*{Acknowledgments}

We are thankful for the open access of the SPARC data set. ZC is supported by the National Natural Science Fund of China under grant Nos. 11675182 and 11690022.


\newpage
\appendices
\section{The best fitting value for full SPARC sample}

\onecolumn
\setlength\LTcapwidth{\textheight}
\begin{landscape}
	\begin{longtable}{cc r@{ $\pm$ }l r@{ $\pm$ }l r@{ $\pm$ }l r@{ $\pm$ }l r@{ $\pm$ }l r@{ $\pm$ }l r@{ $\pm$ }l r@{ $\pm$ }l r@{ $\pm$ }l c}
		\caption{\label{sample}The best-fitting values of galactic parameters and dark halo parameters for the Burkert profile with flat prior. These galaxies are sorted by luminosity. The scale radius $r_0$, central density $\rho_0$, and its product $\rho_0r_0$ are deduced from the best-fitting values of halo parameters. The reduced $\chi^2$ for those galaxies whose data points more than the fitting parameters are listed in the last column.}\\
		\hline
		SPARC ID&Galaxy Name&\multicolumn{2}{c}{$\delta_D$}&\multicolumn{2}{c}{$\delta_i$}&\multicolumn{2}{c}{$\Upsilon_{d}$}&\multicolumn{2}{c}{$\Upsilon_{b}$}&\multicolumn{2}{c}{$\log_{10}V_{200}$}&\multicolumn{2}{c}{$\log_{10}C_{200}$}&\multicolumn{2}{c}{$\log_{10}r_0$}&\multicolumn{2}{c}{$\log_{10}\rho_0$}&\multicolumn{2}{c}{$\log_{10}\rho_0r_0$}&$\chi^2_{\nu}$ \\
		&&\multicolumn{2}{c}{}&\multicolumn{2}{c}{}&\multicolumn{2}{c}{}&\multicolumn{2}{c}{}&\multicolumn{2}{c}{$\mathrm{(km\,s^{-2})}$}&\multicolumn{2}{c}{}&\multicolumn{2}{c}{$\mathrm{(kpc)}$}&\multicolumn{2}{c}{$\mathrm{(M_{\odot}\,pc^{-3})}$}&\multicolumn{2}{c}{$\mathrm{(M_{\odot}\,pc^{-2})}$}& \\
		\hline
		\endfirsthead
		
		\caption{Continued}\\
		\hline
		SPARC ID&Galaxy Name&\multicolumn{2}{c}{$\delta_D$}&\multicolumn{2}{c}{$\delta_i$}&\multicolumn{2}{c}{$\Upsilon_{d}$}&\multicolumn{2}{c}{$\Upsilon_{b}$}&\multicolumn{2}{c}{$\log_{10}V_{200}$}&\multicolumn{2}{c}{$\log_{10}C_{200}$}&\multicolumn{2}{c}{$\log_{10}r_0$}&\multicolumn{2}{c}{$\log_{10}\rho_0$}&\multicolumn{2}{c}{$\log_{10}\rho_0r_0$}&$\chi^2_{\nu}$ \\
		&&\multicolumn{2}{c}{}&\multicolumn{2}{c}{}&\multicolumn{2}{c}{}&\multicolumn{2}{c}{}&\multicolumn{2}{c}{$\mathrm{(km\,s^{-2})}$}&\multicolumn{2}{c}{}&\multicolumn{2}{c}{$\mathrm{(kpc)}$}&\multicolumn{2}{c}{$\mathrm{(M_{\odot}\,pc^{-3})}$}&\multicolumn{2}{c}{$\mathrm{(M_{\odot}\,pc^{-2})}$}& \\
		\hline
		\endhead
		
		001 &     UGC02487 & 1.16 & 0.14 & 1.20 & 0.12 & 0.68 & 0.14 & 0.69 & 0.14 & 2.26 & 0.04 & 1.10 & 0.14 & 1.30 & 0.14 & -1.98 & 0.37 & 2.33 & 0.23 & 4.807 \\
		002 &     UGC02885 & 1.08 & 0.09 & 1.02 & 0.06 & 0.57 & 0.11 & 0.97 & 0.11 & 2.32 & 0.03 & 1.00 & 0.09 & 1.46 & 0.11 & -2.22 & 0.21 & 2.24 & 0.11 & 1.097 \\
		003 &      NGC6195 & 0.99 & 0.09 & 0.99 & 0.07 & 0.40 & 0.08 & 0.78 & 0.09 & 2.24 & 0.10 & 1.01 & 0.11 & 1.36 & 0.20 & -2.18 & 0.27 & 2.18 & 0.10 & 1.789 \\
		004 &     UGC11455 & 0.67 & 0.13 & 1.00 & 0.01 & 0.32 & 0.06 & \multicolumn{2}{c}{......} & 2.17 & 0.02 & 1.59 & 0.08 & 0.71 & 0.10 & -0.71 & 0.21 & 3.01 & 0.11 & 1.699 \\
		005 &      NGC5371 & 0.76 & 0.13 & 1.01 & 0.04 & 0.86 & 0.13 & \multicolumn{2}{c}{......} & 2.69 & 0.05 & 0.63 & 0.07 & 2.19 & 0.09 & -3.06 & 0.15 & 2.13 & 0.09 & 10.263 \\
		006 &      NGC2955 & 0.97 & 0.09 & 0.92 & 0.10 & 0.33 & 0.06 & 0.94 & 0.15 & 2.20 & 0.04 & 1.30 & 0.06 & 1.04 & 0.07 & -1.47 & 0.16 & 2.57 & 0.11 & 3.731 \\
		007 &      NGC0801 & 1.04 & 0.09 & 1.00 & 0.01 & 0.63 & 0.06 & \multicolumn{2}{c}{......} & 2.22 & 0.05 & 0.76 & 0.07 & 1.60 & 0.11 & -2.79 & 0.15 & 1.82 & 0.06 & 8.122 \\
		008 &  ESO563-G021 & 0.55 & 0.14 & 0.99 & 0.04 & 0.30 & 0.07 & \multicolumn{2}{c}{......} & 2.22 & 0.03 & 1.72 & 0.10 & 0.64 & 0.12 & -0.35 & 0.26 & 3.28 & 0.13 & 8.771 \\
		009 &     UGC09133 & 1.22 & 0.15 & 1.07 & 0.09 & 0.73 & 0.10 & 0.40 & 0.05 & 2.22 & 0.02 & 0.84 & 0.05 & 1.51 & 0.06 & -2.59 & 0.11 & 1.92 & 0.06 & 7.886 \\
		010 &     UGC02953 & 0.49 & 0.12 & 0.94 & 0.07 & 0.14 & 0.02 & 1.63 & 0.20 & 2.11 & 0.04 & 1.99 & 0.05 & 0.25 & 0.08 & 0.39 & 0.13 & 3.64 & 0.06 & 7.952 \\
		011 &      NGC7331 & 0.89 & 0.09 & 1.00 & 0.03 & 0.47 & 0.06 & 0.58 & 0.13 & 2.19 & 0.02 & 1.19 & 0.05 & 1.14 & 0.06 & -1.75 & 0.12 & 2.40 & 0.06 & 1.287 \\
		012 &      NGC3992 & 0.98 & 0.10 & 1.00 & 0.04 & 0.45 & 0.13 & \multicolumn{2}{c}{......} & 2.15 & 0.05 & 1.48 & 0.15 & 0.80 & 0.19 & -0.99 & 0.37 & 2.82 & 0.18 & 1.400 \\
		013 &      NGC6674 & 1.37 & 0.16 & 1.15 & 0.09 & 0.85 & 0.15 & 0.73 & 0.16 & 2.41 & 0.05 & 0.48 & 0.08 & 2.06 & 0.11 & -3.37 & 0.15 & 1.69 & 0.07 & 1.440 \\
		014 &      NGC5985 & 0.66 & 0.21 & 0.99 & 0.03 & 0.30 & 0.06 & 0.81 & 0.21 & 2.14 & 0.03 & 1.81 & 0.10 & 0.47 & 0.13 & -0.10 & 0.28 & 3.37 & 0.15 & 2.990 \\
		015 &      NGC2841 & 1.13 & 0.08 & 1.07 & 0.07 & 1.08 & 0.10 & 0.85 & 0.09 & 2.30 & 0.01 & 1.02 & 0.04 & 1.42 & 0.05 & -2.17 & 0.09 & 2.25 & 0.04 & 1.810 \\
		016 &       IC4202 & 0.31 & 0.05 & 1.00 & 0.01 & 0.37 & 0.07 & 0.20 & 0.03 & 2.00 & 0.01 & 2.00 & 0.04 & 0.14 & 0.05 & 0.41 & 0.11 & 3.55 & 0.06 & 4.646 \\
		017 &      NGC5005 & 0.96 & 0.08 & 1.00 & 0.03 & 0.50 & 0.09 & 0.59 & 0.09 & 2.08 & 0.24 & 1.44 & 0.20 & 0.78 & 0.41 & -1.11 & 0.51 & 2.67 & 0.21 & 0.099 \\
		018 &      NGC5907 & 0.98 & 0.05 & 1.00 & 0.02 & 0.19 & 0.03 & \multicolumn{2}{c}{......} & 2.10 & 0.01 & 1.58 & 0.03 & 0.66 & 0.03 & -0.73 & 0.07 & 2.93 & 0.04 & 8.533 \\
		019 &     UGC05253 & 0.85 & 0.15 & 0.96 & 0.09 & 0.35 & 0.07 & 0.93 & 0.16 & 2.14 & 0.04 & 1.45 & 0.06 & 0.82 & 0.08 & -1.06 & 0.15 & 2.77 & 0.09 & 0.956 \\
		020 &      NGC5055 & 1.01 & 0.03 & 1.10 & 0.09 & 0.38 & 0.04 & \multicolumn{2}{c}{......} & 2.06 & 0.02 & 1.16 & 0.03 & 1.03 & 0.02 & -1.81 & 0.07 & 2.22 & 0.06 & 2.202 \\
		021 &      NGC2998 & 1.15 & 0.13 & 1.01 & 0.03 & 0.72 & 0.09 & \multicolumn{2}{c}{......} & 2.16 & 0.04 & 0.90 & 0.08 & 1.40 & 0.12 & -2.45 & 0.20 & 1.94 & 0.09 & 4.002 \\
		022 &     UGC11914 & 1.05 & 0.18 & 1.13 & 0.10 & 0.45 & 0.07 & 0.78 & 0.13 & 2.70 & 0.07 & 1.22 & 0.05 & 1.62 & 0.10 & -1.67 & 0.13 & 2.95 & 0.08 & 0.588 \\
		023 &      NGC3953 & 0.97 & 0.14 & 1.00 & 0.02 & 0.47 & 0.14 & \multicolumn{2}{c}{......} & 1.96 & 0.25 & 1.58 & 0.37 & 0.52 & 0.55 & -0.73 & 0.90 & 2.79 & 0.45 & 0.438 \\
		024 &     UGC12506 & 1.00 & 0.10 & 1.00 & 0.04 & 0.42 & 0.09 & \multicolumn{2}{c}{......} & 2.13 & 0.01 & 1.54 & 0.05 & 0.73 & 0.06 & -0.84 & 0.13 & 2.90 & 0.07 & 0.796 \\
		025 &      NGC0891 & 0.94 & 0.05 & 1.00 & 0.01 & 0.24 & 0.03 & 0.62 & 0.07 & 2.04 & 0.01 & 1.55 & 0.04 & 0.63 & 0.05 & -0.81 & 0.10 & 2.82 & 0.06 & 2.032 \\
		026 &     UGC06614 & 0.98 & 0.10 & 0.93 & 0.13 & 0.50 & 0.12 & 0.69 & 0.15 & 2.23 & 0.05 & 1.05 & 0.11 & 1.31 & 0.12 & -2.09 & 0.27 & 2.22 & 0.17 & 0.095 \\
		027 &     UGC02916 & 0.87 & 0.12 & 0.92 & 0.09 & 0.47 & 0.11 & 0.70 & 0.11 & 2.10 & 0.04 & 1.42 & 0.05 & 0.82 & 0.06 & -1.14 & 0.13 & 2.68 & 0.08 & 10.545 \\
		028 &     UGC03205 & 0.92 & 0.15 & 1.00 & 0.06 & 0.15 & 0.03 & 0.85 & 0.17 & 2.06 & 0.02 & 1.68 & 0.06 & 0.51 & 0.08 & -0.46 & 0.17 & 3.06 & 0.09 & 4.358 \\
		029 &      NGC5033 & 1.00 & 0.16 & 1.00 & 0.01 & 0.45 & 0.09 & 0.51 & 0.10 & 2.09 & 0.01 & 1.41 & 0.07 & 0.81 & 0.08 & -1.17 & 0.17 & 2.65 & 0.09 & 2.509 \\
		030 &      NGC4088 & 0.97 & 0.12 & 1.00 & 0.03 & 0.46 & 0.07 & \multicolumn{2}{c}{......} & 2.46 & 0.22 & 0.88 & 0.14 & 1.72 & 0.34 & -2.51 & 0.34 & 2.21 & 0.15 & 0.744 \\
		031 &      NGC4157 & 0.99 & 0.12 & 1.01 & 0.04 & 0.53 & 0.08 & 0.66 & 0.14 & 2.17 & 0.12 & 1.02 & 0.12 & 1.28 & 0.23 & -2.16 & 0.29 & 2.12 & 0.10 & 0.448 \\
		032 &     UGC03546 & 0.87 & 0.15 & 0.98 & 0.08 & 0.56 & 0.11 & 0.59 & 0.11 & 2.04 & 0.03 & 1.40 & 0.08 & 0.78 & 0.09 & -1.21 & 0.20 & 2.57 & 0.11 & 0.712 \\
		033 &     UGC06787 & 2.24 & 0.25 & 1.07 & 0.04 & 0.89 & 0.12 & 0.19 & 0.03 & 2.44 & 0.02 & 0.65 & 0.05 & 1.92 & 0.06 & -3.02 & 0.11 & 1.90 & 0.06 & 17.972 \\
		034 &      NGC4051 & 0.96 & 0.13 & 0.99 & 0.06 & 0.46 & 0.11 & \multicolumn{2}{c}{......} & 1.80 & 0.26 & 1.46 & 0.33 & 0.47 & 0.54 & -1.03 & 0.81 & 2.44 & 0.37 & 2.451 \\
		035 &      NGC4217 & 0.55 & 0.14 & 1.00 & 0.02 & 0.54 & 0.13 & 0.32 & 0.07 & 1.96 & 0.03 & 1.69 & 0.12 & 0.40 & 0.15 & -0.42 & 0.31 & 2.98 & 0.16 & 1.493 \\
		036 &      NGC3521 & 1.08 & 0.19 & 1.00 & 0.07 & 0.54 & 0.10 & \multicolumn{2}{c}{......} & 2.20 & 0.14 & 1.16 & 0.09 & 1.18 & 0.21 & -1.83 & 0.23 & 2.35 & 0.10 & 0.203 \\
		037 &      NGC2903 & 0.80 & 0.17 & 0.99 & 0.05 & 0.46 & 0.12 & \multicolumn{2}{c}{......} & 1.98 & 0.03 & 1.58 & 0.13 & 0.54 & 0.16 & -0.72 & 0.36 & 2.82 & 0.20 & 7.458 \\
		038 &      NGC2683 & 1.00 & 0.05 & 0.99 & 0.06 & 0.49 & 0.10 & 0.73 & 0.16 & 1.96 & 0.02 & 1.51 & 0.12 & 0.58 & 0.13 & -0.91 & 0.32 & 2.67 & 0.19 & 1.245 \\
		039 &      NGC4013 & 1.09 & 0.11 & 1.00 & 0.01 & 0.57 & 0.08 & 0.82 & 0.18 & 2.21 & 0.04 & 0.93 & 0.05 & 1.41 & 0.08 & -2.38 & 0.12 & 2.03 & 0.05 & 0.836 \\
		040 &      NGC7814 & 1.00 & 0.04 & 1.00 & 0.01 & 0.54 & 0.13 & 0.62 & 0.05 & 2.06 & 0.01 & 1.47 & 0.05 & 0.73 & 0.06 & -1.03 & 0.12 & 2.71 & 0.06 & 0.867 \\
		041 &     UGC06786 & 1.22 & 0.16 & 1.02 & 0.04 & 0.66 & 0.09 & 0.72 & 0.10 & 2.17 & 0.02 & 1.23 & 0.07 & 1.07 & 0.08 & -1.65 & 0.17 & 2.43 & 0.09 & 2.269 \\
		042 &      NGC3877 & 0.76 & 0.14 & 1.00 & 0.01 & 0.30 & 0.06 & \multicolumn{2}{c}{......} & 1.89 & 0.02 & 1.68 & 0.07 & 0.35 & 0.09 & -0.46 & 0.20 & 2.88 & 0.11 & 2.727 \\
		043 &      NGC0289 & 1.27 & 0.19 & 1.10 & 0.09 & 0.55 & 0.09 & \multicolumn{2}{c}{......} & 2.06 & 0.03 & 0.90 & 0.09 & 1.30 & 0.09 & -2.46 & 0.21 & 1.84 & 0.12 & 2.144 \\
		044 &      NGC1090 & 0.65 & 0.21 & 1.00 & 0.05 & 0.35 & 0.09 & \multicolumn{2}{c}{......} & 1.95 & 0.04 & 1.60 & 0.15 & 0.48 & 0.19 & -0.66 & 0.41 & 2.82 & 0.22 & 1.478 \\
		045 &      NGC3726 & 1.06 & 0.12 & 1.00 & 0.04 & 0.61 & 0.08 & \multicolumn{2}{c}{......} & 2.31 & 0.18 & 0.80 & 0.11 & 1.64 & 0.28 & -2.69 & 0.25 & 1.96 & 0.12 & 2.935 \\
		046 &     UGC09037 & 0.87 & 0.10 & 0.93 & 0.08 & 0.30 & 0.05 & \multicolumn{2}{c}{......} & 2.03 & 0.03 & 1.25 & 0.07 & 0.92 & 0.08 & -1.59 & 0.17 & 2.33 & 0.10 & 1.091 \\
		047 &      NGC6946 & 0.92 & 0.17 & 0.99 & 0.05 & 0.60 & 0.09 & 0.64 & 0.12 & 1.96 & 0.04 & 1.28 & 0.12 & 0.81 & 0.16 & -1.50 & 0.32 & 2.31 & 0.16 & 1.526 \\
		048 &      NGC4100 & 0.96 & 0.13 & 1.00 & 0.03 & 0.43 & 0.07 & \multicolumn{2}{c}{......} & 1.96 & 0.02 & 1.58 & 0.07 & 0.51 & 0.08 & -0.73 & 0.19 & 2.79 & 0.11 & 0.631 \\
		049 &      NGC3893 & 0.97 & 0.13 & 0.99 & 0.04 & 0.45 & 0.08 & \multicolumn{2}{c}{......} & 1.97 & 0.03 & 1.51 & 0.11 & 0.60 & 0.14 & -0.92 & 0.29 & 2.68 & 0.16 & 0.382 \\
		050 &     UGC06973 & 0.60 & 0.11 & 0.98 & 0.04 & 0.31 & 0.06 & 0.61 & 0.13 & 1.95 & 0.05 & 1.70 & 0.09 & 0.38 & 0.14 & -0.40 & 0.24 & 2.98 & 0.11 & 1.622 \\
		051 &  ESO079-G014 & 0.81 & 0.21 & 1.00 & 0.06 & 0.46 & 0.10 & \multicolumn{2}{c}{......} & 2.05 & 0.02 & 1.38 & 0.11 & 0.80 & 0.12 & -1.24 & 0.28 & 2.56 & 0.15 & 1.267 \\
		052 &     UGC08699 & 1.15 & 0.15 & 1.03 & 0.10 & 0.86 & 0.14 & 0.57 & 0.08 & 2.07 & 0.05 & 1.13 & 0.08 & 1.08 & 0.12 & -1.90 & 0.20 & 2.17 & 0.09 & 0.962 \\
		053 &      NGC4138 & 0.99 & 0.13 & 0.98 & 0.06 & 0.50 & 0.12 & 0.66 & 0.15 & 1.87 & 0.09 & 1.65 & 0.21 & 0.35 & 0.29 & -0.53 & 0.56 & 2.82 & 0.29 & 3.291 \\
		054 &      NGC3198 & 1.06 & 0.09 & 1.00 & 0.04 & 0.66 & 0.07 & \multicolumn{2}{c}{......} & 2.01 & 0.01 & 1.15 & 0.04 & 1.00 & 0.05 & -1.85 & 0.11 & 2.15 & 0.06 & 1.515 \\
		055 &      NGC3949 & 1.00 & 0.12 & 1.00 & 0.04 & 0.53 & 0.08 & \multicolumn{2}{c}{......} & 2.67 & 0.29 & 1.18 & 0.16 & 1.62 & 0.42 & -1.76 & 0.41 & 2.86 & 0.23 & 1.685 \\
		056 &      NGC6015 & 1.62 & 0.26 & 1.01 & 0.04 & 0.66 & 0.15 & \multicolumn{2}{c}{......} & 2.17 & 0.03 & 0.73 & 0.09 & 1.58 & 0.11 & -2.86 & 0.21 & 1.73 & 0.10 & 9.552 \\
		057 &      NGC3917 & 0.95 & 0.14 & 1.00 & 0.03 & 0.44 & 0.10 & \multicolumn{2}{c}{......} & 1.91 & 0.02 & 1.40 & 0.07 & 0.65 & 0.07 & -1.21 & 0.17 & 2.44 & 0.10 & 1.424 \\
		058 &      NGC4085 & 0.86 & 0.13 & 1.00 & 0.02 & 0.35 & 0.07 & \multicolumn{2}{c}{......} & 1.92 & 0.24 & 1.45 & 0.13 & 0.61 & 0.35 & -1.06 & 0.34 & 2.55 & 0.15 & 10.633 \\
		059 &      NGC4389 & 0.76 & 0.14 & 0.90 & 0.08 & 0.29 & 0.06 & \multicolumn{2}{c}{......} & 2.69 & 0.20 & 1.28 & 0.10 & 1.55 & 0.23 & -1.52 & 0.25 & 3.03 & 0.22 & ... \\
		060 &      NGC4559 & 1.02 & 0.21 & 1.00 & 0.01 & 0.53 & 0.10 & \multicolumn{2}{c}{......} & 1.91 & 0.03 & 1.15 & 0.12 & 0.90 & 0.14 & -1.85 & 0.29 & 2.05 & 0.15 & 0.266 \\
		061 &      NGC3769 & 0.97 & 0.13 & 1.00 & 0.03 & 0.45 & 0.08 & \multicolumn{2}{c}{......} & 1.90 & 0.03 & 1.28 & 0.09 & 0.76 & 0.11 & -1.51 & 0.23 & 2.25 & 0.12 & 0.892 \\
		062 &      NGC4010 & 0.97 & 0.14 & 1.00 & 0.01 & 0.45 & 0.10 & \multicolumn{2}{c}{......} & 1.93 & 0.12 & 1.28 & 0.11 & 0.79 & 0.21 & -1.52 & 0.27 & 2.27 & 0.11 & 2.311 \\
		063 &      NGC3972 & 0.97 & 0.14 & 1.00 & 0.01 & 0.44 & 0.12 & \multicolumn{2}{c}{......} & 1.87 & 0.09 & 1.45 & 0.11 & 0.56 & 0.18 & -1.08 & 0.27 & 2.48 & 0.12 & 1.384 \\
		064 &     UGC03580 & 0.59 & 0.11 & 0.98 & 0.06 & 1.10 & 0.18 & 0.29 & 0.05 & 1.93 & 0.03 & 1.32 & 0.06 & 0.75 & 0.08 & -1.42 & 0.15 & 2.33 & 0.07 & 2.680 \\
		065 &      NGC6503 & 0.99 & 0.05 & 1.00 & 0.03 & 0.57 & 0.04 & \multicolumn{2}{c}{......} & 1.87 & 0.01 & 1.31 & 0.02 & 0.69 & 0.03 & -1.43 & 0.06 & 2.26 & 0.03 & 2.631 \\
		066 &     UGC11557 & 0.92 & 0.26 & 0.66 & 0.29 & 0.46 & 0.11 & \multicolumn{2}{c}{......} & 1.93 & 0.25 & 1.34 & 0.34 & 0.72 & 0.47 & -1.37 & 0.87 & 2.36 & 0.53 & 0.771 \\
		067 &     UGC00128 & 0.99 & 0.15 & 1.03 & 0.14 & 0.32 & 0.08 & \multicolumn{2}{c}{......} & 1.96 & 0.04 & 1.20 & 0.05 & 0.90 & 0.06 & -1.73 & 0.14 & 2.18 & 0.10 & 9.834 \\
		068 &      F579-V1 & 1.00 & 0.10 & 1.37 & 0.40 & 0.50 & 0.15 & \multicolumn{2}{c}{......} & 1.59 & 0.25 & 1.60 & 0.30 & 0.12 & 0.46 & -0.67 & 0.71 & 2.46 & 0.40 & 0.197 \\
		069 &      NGC4183 & 0.96 & 0.15 & 1.00 & 0.02 & 0.48 & 0.22 & \multicolumn{2}{c}{......} & 1.80 & 0.14 & 1.42 & 0.20 & 0.52 & 0.33 & -1.14 & 0.48 & 2.38 & 0.18 & 0.464 \\
		070 &       F571-8 & 0.47 & 0.14 & 0.99 & 0.05 & 0.36 & 0.09 & \multicolumn{2}{c}{......} & 1.94 & 0.04 & 1.59 & 0.10 & 0.48 & 0.14 & -0.69 & 0.28 & 2.79 & 0.14 & 0.562 \\
		071 &      NGC2403 & 0.94 & 0.05 & 0.93 & 0.05 & 0.90 & 0.07 & \multicolumn{2}{c}{......} & 1.93 & 0.01 & 1.32 & 0.02 & 0.74 & 0.02 & -1.40 & 0.05 & 2.34 & 0.04 & 11.803 \\
		072 &     UGC06930 & 1.00 & 0.14 & 1.00 & 0.17 & 0.49 & 0.14 & \multicolumn{2}{c}{......} & 1.79 & 0.10 & 1.37 & 0.15 & 0.56 & 0.17 & -1.27 & 0.37 & 2.29 & 0.25 & 0.354 \\
		073 &       F568-3 & 1.00 & 0.10 & 0.83 & 0.23 & 0.48 & 0.11 & \multicolumn{2}{c}{......} & 1.96 & 0.11 & 1.28 & 0.12 & 0.82 & 0.07 & -1.51 & 0.32 & 2.31 & 0.29 & 1.289 \\
		074 &     UGC01230 & 0.98 & 0.20 & 0.85 & 0.38 & 0.49 & 0.12 & \multicolumn{2}{c}{......} & 1.90 & 0.22 & 1.44 & 0.21 & 0.60 & 0.12 & -1.10 & 0.56 & 2.50 & 0.54 & 0.434 \\
		075 &      NGC0247 & 1.06 & 0.05 & 1.03 & 0.04 & 2.12 & 0.60 & \multicolumn{2}{c}{......} & 2.66 & 0.42 & 0.60 & 0.34 & 2.19 & 0.75 & -3.13 & 0.81 & 2.06 & 0.18 & 2.178 \\
		076 &      NGC7793 & 1.01 & 0.05 & 1.18 & 0.14 & 0.58 & 0.09 & \multicolumn{2}{c}{......} & 1.67 & 0.23 & 1.28 & 0.14 & 0.53 & 0.35 & -1.51 & 0.36 & 2.01 & 0.16 & 0.784 \\
		077 &     UGC06917 & 0.99 & 0.14 & 0.99 & 0.04 & 0.49 & 0.11 & \multicolumn{2}{c}{......} & 1.81 & 0.03 & 1.39 & 0.07 & 0.55 & 0.09 & -1.23 & 0.19 & 2.33 & 0.10 & 0.346 \\
		078 &      NGC1003 & 1.29 & 0.20 & 1.03 & 0.07 & 0.74 & 0.12 & \multicolumn{2}{c}{......} & 1.98 & 0.02 & 0.86 & 0.07 & 1.26 & 0.08 & -2.55 & 0.17 & 1.70 & 0.09 & 3.117 \\
		079 &       F574-1 & 0.99 & 0.10 & 0.98 & 0.16 & 0.49 & 0.12 & \multicolumn{2}{c}{......} & 1.79 & 0.04 & 1.39 & 0.06 & 0.54 & 0.06 & -1.23 & 0.16 & 2.31 & 0.12 & 0.134 \\
		080 &       F568-1 & 0.99 & 0.11 & 0.99 & 0.20 & 0.49 & 0.12 & \multicolumn{2}{c}{......} & 1.91 & 0.09 & 1.46 & 0.09 & 0.59 & 0.08 & -1.05 & 0.24 & 2.54 & 0.21 & 0.149 \\
		081 &     UGC06983 & 1.01 & 0.14 & 1.00 & 0.02 & 0.51 & 0.12 & \multicolumn{2}{c}{......} & 1.80 & 0.02 & 1.43 & 0.07 & 0.51 & 0.08 & -1.12 & 0.18 & 2.39 & 0.10 & 0.575 \\
		082 &     UGC05986 & 1.08 & 0.26 & 1.00 & 0.02 & 0.50 & 0.11 & \multicolumn{2}{c}{......} & 1.85 & 0.03 & 1.43 & 0.11 & 0.56 & 0.14 & -1.13 & 0.29 & 2.43 & 0.15 & 1.244 \\
		083 &      NGC0055 & 0.99 & 0.05 & 1.00 & 0.04 & 0.39 & 0.07 & \multicolumn{2}{c}{......} & 1.81 & 0.02 & 1.14 & 0.03 & 0.80 & 0.04 & -1.86 & 0.08 & 1.94 & 0.04 & 0.415 \\
		084 &  ESO116-G012 & 1.06 & 0.25 & 1.00 & 0.04 & 0.53 & 0.11 & \multicolumn{2}{c}{......} & 1.84 & 0.03 & 1.37 & 0.11 & 0.60 & 0.14 & -1.27 & 0.30 & 2.33 & 0.16 & 1.090 \\
		085 &     UGC07323 & 1.08 & 0.27 & 1.01 & 0.06 & 0.51 & 0.12 & \multicolumn{2}{c}{......} & 1.81 & 0.32 & 1.19 & 0.18 & 0.76 & 0.46 & -1.75 & 0.45 & 2.01 & 0.25 & 0.518 \\
		086 &     UGC05005 & 1.02 & 0.21 & 0.98 & 0.26 & 0.49 & 0.12 & \multicolumn{2}{c}{......} & 1.91 & 0.15 & 1.01 & 0.23 & 1.03 & 0.31 & -2.19 & 0.54 & 1.84 & 0.31 & 0.024 \\
		087 &       F561-1 & 1.00 & 0.15 & 0.79 & 0.32 & 0.48 & 0.12 & \multicolumn{2}{c}{......} & 1.54 & 0.36 & 1.29 & 0.38 & 0.39 & 0.59 & -1.49 & 0.93 & 1.90 & 0.59 & ... \\
		088 &      NGC0024 & 0.99 & 0.06 & 0.99 & 0.06 & 0.44 & 0.65 & \multicolumn{2}{c}{......} & 1.75 & 0.24 & 1.71 & 0.34 & 0.18 & 0.56 & -0.38 & 0.87 & 2.80 & 0.35 & 0.893 \\
		089 &      F568-V1 & 1.00 & 0.10 & 1.01 & 0.27 & 0.49 & 0.12 & \multicolumn{2}{c}{......} & 1.83 & 0.11 & 1.50 & 0.12 & 0.47 & 0.08 & -0.94 & 0.31 & 2.53 & 0.28 & 0.079 \\
		090 &     UGC06628 & 0.97 & 0.29 & 0.75 & 0.30 & 0.49 & 0.12 & \multicolumn{2}{c}{......} & 1.41 & 0.43 & 1.53 & 0.51 & 0.01 & 0.85 & -0.85 & 1.24 & 2.16 & 0.60 & 0.246 \\
		091 &     UGC02455 & 0.19 & 0.10 & 0.98 & 0.10 & 0.46 & 0.11 & \multicolumn{2}{c}{......} & 2.68 & 0.23 & 1.65 & 0.14 & 1.16 & 0.29 & -0.53 & 0.39 & 3.63 & 0.31 & 1.647 \\
		092 &     UGC07089 & 1.01 & 0.14 & 0.99 & 0.04 & 0.51 & 0.14 & \multicolumn{2}{c}{......} & 1.78 & 0.31 & 1.10 & 0.13 & 0.81 & 0.43 & -1.96 & 0.33 & 1.85 & 0.20 & 0.252 \\
		093 &     UGC05999 & 0.96 & 0.21 & 0.79 & 0.35 & 0.49 & 0.12 & \multicolumn{2}{c}{......} & 1.95 & 0.24 & 1.24 & 0.24 & 0.85 & 0.26 & -1.61 & 0.63 & 2.24 & 0.54 & ... \\
		094 &      NGC2976 & 0.99 & 0.05 & 1.00 & 0.14 & 0.49 & 0.10 & \multicolumn{2}{c}{......} & 1.83 & 0.30 & 1.45 & 0.10 & 0.51 & 0.37 & -1.06 & 0.26 & 2.46 & 0.26 & 0.361 \\
		095 &     UGC05750 & 0.98 & 0.21 & 1.00 & 0.17 & 0.49 & 0.13 & \multicolumn{2}{c}{......} & 1.79 & 0.12 & 1.06 & 0.14 & 0.87 & 0.21 & -2.07 & 0.34 & 1.80 & 0.21 & 0.210 \\
		096 &      NGC0100 & 0.98 & 0.29 & 1.00 & 0.01 & 0.49 & 0.11 & \multicolumn{2}{c}{......} & 1.80 & 0.09 & 1.27 & 0.17 & 0.66 & 0.24 & -1.53 & 0.43 & 2.13 & 0.20 & 0.107 \\
		097 &     UGC00634 & 0.99 & 0.25 & 1.03 & 0.22 & 0.49 & 0.12 & \multicolumn{2}{c}{......} & 1.89 & 0.08 & 1.19 & 0.16 & 0.84 & 0.15 & -1.75 & 0.40 & 2.08 & 0.28 & ... \\
		098 &      F563-V2 & 1.00 & 0.20 & 0.95 & 0.34 & 0.52 & 0.12 & \multicolumn{2}{c}{......} & 1.85 & 0.16 & 1.54 & 0.16 & 0.45 & 0.12 & -0.83 & 0.43 & 2.62 & 0.40 & 0.353 \\
		099 &      NGC5585 & 1.15 & 0.20 & 1.00 & 0.04 & 0.53 & 0.10 & \multicolumn{2}{c}{......} & 1.82 & 0.02 & 1.18 & 0.06 & 0.78 & 0.07 & -1.78 & 0.16 & 2.00 & 0.09 & 5.256 \\
		100 &      NGC0300 & 1.01 & 0.05 & 1.11 & 0.23 & 0.56 & 0.14 & \multicolumn{2}{c}{......} & 1.76 & 0.07 & 1.28 & 0.12 & 0.62 & 0.12 & -1.51 & 0.32 & 2.11 & 0.23 & 0.656 \\
		101 &     UGC06923 & 1.00 & 0.14 & 1.00 & 0.03 & 0.47 & 0.12 & \multicolumn{2}{c}{......} & 1.73 & 0.28 & 1.34 & 0.14 & 0.52 & 0.41 & -1.35 & 0.36 & 2.17 & 0.18 & ... \\
		102 &       F574-2 & 0.99 & 0.10 & 0.96 & 0.24 & 0.51 & 0.12 & \multicolumn{2}{c}{......} & 1.30 & 0.53 & 0.63 & 0.46 & 0.81 & 0.88 & -3.07 & 1.06 & 0.74 & 0.50 & ... \\
		103 &     UGC07125 & 0.84 & 0.21 & 1.00 & 0.02 & 0.50 & 0.12 & \multicolumn{2}{c}{......} & 1.62 & 0.02 & 1.15 & 0.12 & 0.61 & 0.12 & -1.84 & 0.29 & 1.77 & 0.17 & 0.290 \\
		104 &     UGC07524 & 1.00 & 0.05 & 1.01 & 0.06 & 0.54 & 0.12 & \multicolumn{2}{c}{......} & 1.70 & 0.03 & 1.28 & 0.03 & 0.56 & 0.03 & -1.52 & 0.08 & 2.04 & 0.07 & 0.260 \\
		105 &     UGC06399 & 0.99 & 0.14 & 1.00 & 0.03 & 0.50 & 0.12 & \multicolumn{2}{c}{......} & 1.75 & 0.04 & 1.35 & 0.07 & 0.54 & 0.10 & -1.33 & 0.19 & 2.21 & 0.09 & 0.110 \\
		106 &     UGC07151 & 1.00 & 0.05 & 1.00 & 0.02 & 0.52 & 0.22 & \multicolumn{2}{c}{......} & 1.57 & 0.21 & 1.51 & 0.14 & 0.20 & 0.34 & -0.92 & 0.36 & 2.28 & 0.09 & 1.627 \\
		107 &       F567-2 & 1.01 & 0.15 & 0.93 & 0.46 & 0.49 & 0.13 & \multicolumn{2}{c}{......} & 1.54 & 0.36 & 1.27 & 0.39 & 0.40 & 0.55 & -1.53 & 0.94 & 1.87 & 0.66 & ... \\
		108 &     UGC04325 & 0.92 & 0.28 & 1.00 & 0.07 & 0.50 & 0.12 & \multicolumn{2}{c}{......} & 1.65 & 0.04 & 1.69 & 0.11 & 0.10 & 0.13 & -0.43 & 0.30 & 2.67 & 0.17 & 1.192 \\
		109 &     UGC00191 & 1.13 & 0.35 & 1.05 & 0.13 & 0.46 & 0.26 & \multicolumn{2}{c}{......} & 1.64 & 0.33 & 1.42 & 0.44 & 0.36 & 0.75 & -1.15 & 1.06 & 2.21 & 0.39 & 9.026 \\
		110 &       F563-1 & 0.99 & 0.20 & 0.97 & 0.22 & 0.49 & 0.12 & \multicolumn{2}{c}{......} & 1.86 & 0.10 & 1.39 & 0.12 & 0.60 & 0.11 & -1.23 & 0.31 & 2.38 & 0.25 & 0.765 \\
		111 &      F571-V1 & 1.00 & 0.10 & 1.01 & 0.37 & 0.48 & 0.12 & \multicolumn{2}{c}{......} & 1.78 & 0.20 & 1.18 & 0.24 & 0.73 & 0.31 & -1.76 & 0.60 & 1.97 & 0.42 & 0.164 \\
		112 &     UGC07261 & 0.98 & 0.30 & 0.93 & 0.35 & 0.49 & 0.12 & \multicolumn{2}{c}{......} & 1.62 & 0.29 & 1.58 & 0.37 & 0.18 & 0.59 & -0.74 & 0.93 & 2.44 & 0.47 & 1.783 \\
		113 &     UGC10310 & 0.94 & 0.31 & 1.00 & 0.19 & 0.50 & 0.14 & \multicolumn{2}{c}{......} & 1.61 & 0.16 & 1.44 & 0.23 & 0.31 & 0.33 & -1.09 & 0.56 & 2.21 & 0.31 & 0.445 \\
		114 &     UGC02259 & 1.02 & 0.29 & 1.01 & 0.08 & 0.46 & 0.11 & \multicolumn{2}{c}{......} & 1.66 & 0.04 & 1.59 & 0.11 & 0.21 & 0.13 & -0.70 & 0.30 & 2.51 & 0.17 & 6.781 \\
		115 &       F583-4 & 1.00 & 0.21 & 1.04 & 0.20 & 0.51 & 0.16 & \multicolumn{2}{c}{......} & 1.60 & 0.26 & 1.32 & 0.23 & 0.41 & 0.45 & -1.40 & 0.57 & 2.01 & 0.25 & 0.740 \\
		116 &     UGC12732 & 1.48 & 0.27 & 1.23 & 0.13 & 0.53 & 0.18 & \multicolumn{2}{c}{......} & 1.68 & 0.10 & 1.09 & 0.15 & 0.73 & 0.22 & -2.00 & 0.37 & 1.73 & 0.20 & 0.527 \\
		117 &     UGC06818 & 0.95 & 0.14 & 0.99 & 0.04 & 0.46 & 0.09 & \multicolumn{2}{c}{......} & 2.11 & 0.24 & 1.08 & 0.07 & 1.16 & 0.28 & -2.01 & 0.17 & 2.15 & 0.22 & 3.257 \\
		118 &     UGC04499 & 0.92 & 0.32 & 1.00 & 0.06 & 0.49 & 0.13 & \multicolumn{2}{c}{......} & 1.65 & 0.18 & 1.35 & 0.21 & 0.44 & 0.36 & -1.34 & 0.52 & 2.10 & 0.21 & 0.514 \\
		119 &      F563-V1 & 1.02 & 0.18 & 0.99 & 0.16 & 0.51 & 0.12 & \multicolumn{2}{c}{......} & 1.12 & 0.60 & 0.97 & 0.45 & 0.28 & 0.99 & -2.28 & 1.01 & 1.00 & 0.41 & ... \\
		120 &     UGC06667 & 1.01 & 0.14 & 1.00 & 0.01 & 0.49 & 0.12 & \multicolumn{2}{c}{......} & 1.75 & 0.02 & 1.39 & 0.05 & 0.50 & 0.07 & -1.23 & 0.14 & 2.26 & 0.07 & 0.221 \\
		121 &     UGC02023 & 0.99 & 0.29 & 0.88 & 0.41 & 0.50 & 0.12 & \multicolumn{2}{c}{......} & 2.60 & 0.31 & 1.14 & 0.34 & 1.59 & 0.45 & -1.86 & 0.84 & 2.73 & 0.60 & ... \\
		122 &     UGC04278 & 1.17 & 0.22 & 0.99 & 0.02 & 0.61 & 0.15 & \multicolumn{2}{c}{......} & 1.95 & 0.19 & 1.15 & 0.09 & 0.94 & 0.25 & -1.85 & 0.23 & 2.08 & 0.15 & 0.629 \\
		123 &     UGC12632 & 1.07 & 0.29 & 1.00 & 0.07 & 0.51 & 0.12 & \multicolumn{2}{c}{......} & 1.65 & 0.03 & 1.31 & 0.12 & 0.48 & 0.14 & -1.44 & 0.30 & 2.04 & 0.18 & 0.090 \\
		124 &     UGC08286 & 1.00 & 0.03 & 1.00 & 0.02 & 0.42 & 0.09 & \multicolumn{2}{c}{......} & 1.67 & 0.01 & 1.55 & 0.02 & 0.25 & 0.03 & -0.79 & 0.05 & 2.46 & 0.03 & 1.588 \\
		125 &     UGC07399 & 1.13 & 0.29 & 1.01 & 0.05 & 0.54 & 0.13 & \multicolumn{2}{c}{......} & 1.72 & 0.04 & 1.61 & 0.11 & 0.25 & 0.14 & -0.65 & 0.30 & 2.60 & 0.16 & 1.614 \\
		126 &      NGC4214 & 1.01 & 0.05 & 1.48 & 0.22 & 0.55 & 0.12 & \multicolumn{2}{c}{......} & 1.61 & 0.24 & 1.25 & 0.16 & 0.49 & 0.36 & -1.59 & 0.41 & 1.90 & 0.22 & 1.205 \\
		127 &     UGC05414 & 1.01 & 0.32 & 1.00 & 0.05 & 0.49 & 0.13 & \multicolumn{2}{c}{......} & 1.64 & 0.36 & 1.28 & 0.21 & 0.49 & 0.54 & -1.52 & 0.53 & 1.98 & 0.24 & ... \\
		128 &     UGC08490 & 1.09 & 0.11 & 1.04 & 0.06 & 0.81 & 0.19 & \multicolumn{2}{c}{......} & 1.64 & 0.02 & 1.48 & 0.09 & 0.30 & 0.10 & -0.98 & 0.24 & 2.31 & 0.14 & 0.431 \\
		129 &       IC2574 & 1.03 & 0.05 & 1.05 & 0.08 & 0.71 & 0.10 & \multicolumn{2}{c}{......} & 1.95 & 0.15 & 0.87 & 0.04 & 1.22 & 0.18 & -2.53 & 0.09 & 1.69 & 0.12 & 2.482 \\
		130 &     UGC06446 & 1.17 & 0.29 & 1.01 & 0.06 & 0.50 & 0.13 & \multicolumn{2}{c}{......} & 1.67 & 0.03 & 1.45 & 0.11 & 0.35 & 0.13 & -1.07 & 0.28 & 2.29 & 0.16 & 0.552 \\
		131 &       F583-1 & 0.91 & 0.24 & 0.98 & 0.08 & 0.48 & 0.12 & \multicolumn{2}{c}{......} & 1.77 & 0.04 & 1.29 & 0.10 & 0.62 & 0.13 & -1.49 & 0.27 & 2.13 & 0.15 & 0.218 \\
		132 &     UGC11820 & 1.36 & 0.23 & 1.22 & 0.17 & 1.38 & 0.18 & \multicolumn{2}{c}{......} & 1.81 & 0.27 & 0.68 & 0.22 & 1.26 & 0.47 & -2.96 & 0.47 & 1.30 & 0.18 & 3.192 \\
		133 &     UGC07690 & 0.99 & 0.29 & 1.00 & 0.13 & 0.50 & 0.13 & \multicolumn{2}{c}{......} & 1.41 & 0.28 & 1.70 & 0.32 & -0.15 & 0.57 & -0.40 & 0.83 & 2.45 & 0.35 & 0.573 \\
		134 &     UGC04305 & 1.00 & 0.05 & 0.55 & 0.19 & 0.52 & 0.18 & \multicolumn{2}{c}{......} & 1.47 & 0.32 & 1.58 & 0.45 & 0.02 & 0.72 & -0.71 & 1.04 & 2.31 & 0.44 & 1.142 \\
		135 &      NGC2915 & 0.99 & 0.05 & 0.97 & 0.07 & 0.42 & 0.09 & \multicolumn{2}{c}{......} & 1.70 & 0.02 & 1.53 & 0.05 & 0.31 & 0.06 & -0.85 & 0.12 & 2.46 & 0.07 & 0.577 \\
		136 &     UGC05716 & 1.45 & 0.20 & 1.26 & 0.14 & 0.76 & 0.17 & \multicolumn{2}{c}{......} & 1.64 & 0.03 & 1.05 & 0.07 & 0.72 & 0.07 & -2.10 & 0.18 & 1.62 & 0.12 & 2.695 \\
		137 &     UGC05829 & 1.06 & 0.31 & 1.10 & 0.32 & 0.49 & 0.15 & \multicolumn{2}{c}{......} & 1.56 & 0.28 & 1.20 & 0.22 & 0.50 & 0.40 & -1.72 & 0.54 & 1.77 & 0.40 & 0.327 \\
		138 &      F565-V2 & 0.99 & 0.20 & 0.99 & 0.17 & 0.49 & 0.12 & \multicolumn{2}{c}{......} & 1.82 & 0.17 & 1.22 & 0.13 & 0.74 & 0.26 & -1.68 & 0.32 & 2.06 & 0.18 & 0.308 \\
		139 &       DDO161 & 1.02 & 0.22 & 1.01 & 0.12 & 0.50 & 0.10 & \multicolumn{2}{c}{......} & 1.74 & 0.04 & 0.99 & 0.11 & 0.88 & 0.14 & -2.23 & 0.28 & 1.65 & 0.14 & 0.305 \\
		140 &       DDO170 & 0.70 & 0.26 & 0.98 & 0.11 & 0.48 & 0.12 & \multicolumn{2}{c}{......} & 1.62 & 0.04 & 1.30 & 0.14 & 0.45 & 0.17 & -1.46 & 0.36 & 1.99 & 0.20 & 3.953 \\
		141 &      NGC1705 & 1.00 & 0.05 & 1.02 & 0.10 & 0.56 & 0.14 & \multicolumn{2}{c}{......} & 1.53 & 0.03 & 1.78 & 0.07 & -0.11 & 0.09 & -0.19 & 0.20 & 2.71 & 0.11 & 0.925 \\
		142 &     UGC05721 & 1.08 & 0.26 & 1.01 & 0.08 & 0.52 & 0.12 & \multicolumn{2}{c}{......} & 1.61 & 0.03 & 1.69 & 0.11 & 0.05 & 0.13 & -0.42 & 0.28 & 2.63 & 0.16 & 0.467 \\
		143 &     UGC08837 & 1.00 & 0.05 & 1.00 & 0.06 & 0.47 & 0.09 & \multicolumn{2}{c}{......} & 2.61 & 0.25 & 0.98 & 0.04 & 1.77 & 0.27 & -2.26 & 0.09 & 2.50 & 0.24 & 1.007 \\
		144 &     UGC07603 & 0.97 & 0.29 & 1.00 & 0.04 & 0.48 & 0.11 & \multicolumn{2}{c}{......} & 1.56 & 0.05 & 1.54 & 0.14 & 0.16 & 0.18 & -0.84 & 0.37 & 2.32 & 0.19 & 0.328 \\
		145 &     UGC00891 & 0.95 & 0.29 & 1.00 & 0.09 & 0.49 & 0.12 & \multicolumn{2}{c}{......} & 1.73 & 0.05 & 1.16 & 0.15 & 0.70 & 0.20 & -1.81 & 0.37 & 1.89 & 0.18 & ... \\
		146 &     UGC01281 & 0.98 & 0.04 & 1.00 & 0.01 & 0.50 & 0.12 & \multicolumn{2}{c}{......} & 1.62 & 0.04 & 1.29 & 0.03 & 0.47 & 0.07 & -1.49 & 0.09 & 1.99 & 0.04 & 0.347 \\
		147 &     UGC09992 & 1.00 & 0.27 & 0.99 & 0.31 & 0.48 & 0.13 & \multicolumn{2}{c}{......} & 1.18 & 0.58 & 1.59 & 0.61 & -0.27 & 1.14 & -0.71 & 1.47 & 2.03 & 0.52 & ... \\
		148 &       D512-2 & 1.03 & 0.32 & 1.03 & 0.19 & 0.51 & 0.14 & \multicolumn{2}{c}{......} & 1.37 & 0.33 & 1.33 & 0.23 & 0.18 & 0.51 & -1.39 & 0.60 & 1.79 & 0.30 & ... \\
		149 &     UGC00731 & 1.25 & 0.26 & 1.01 & 0.05 & 0.52 & 0.12 & \multicolumn{2}{c}{......} & 1.64 & 0.02 & 1.33 & 0.07 & 0.45 & 0.08 & -1.39 & 0.18 & 2.06 & 0.11 & 0.658 \\
		150 &     UGC08550 & 1.18 & 0.28 & 1.00 & 0.02 & 0.51 & 0.15 & \multicolumn{2}{c}{......} & 1.51 & 0.04 & 1.45 & 0.12 & 0.20 & 0.16 & -1.06 & 0.32 & 2.14 & 0.17 & 1.548 \\
		151 &     UGC07608 & 1.00 & 0.29 & 0.96 & 0.40 & 0.49 & 0.12 & \multicolumn{2}{c}{......} & 1.71 & 0.28 & 1.36 & 0.23 & 0.48 & 0.36 & -1.30 & 0.60 & 2.18 & 0.47 & 0.240 \\
		152 &      NGC4068 & 1.00 & 0.05 & 0.99 & 0.14 & 0.49 & 0.11 & \multicolumn{2}{c}{......} & 2.47 & 0.35 & 1.12 & 0.10 & 1.49 & 0.39 & -1.91 & 0.24 & 2.57 & 0.32 & ... \\
		153 &      NGC2366 & 1.00 & 0.05 & 0.99 & 0.07 & 0.38 & 0.07 & \multicolumn{2}{c}{......} & 1.54 & 0.02 & 1.29 & 0.03 & 0.39 & 0.04 & -1.49 & 0.09 & 1.90 & 0.06 & 0.908 \\
		154 &     UGC05918 & 0.99 & 0.31 & 1.00 & 0.11 & 0.51 & 0.12 & \multicolumn{2}{c}{......} & 1.42 & 0.07 & 1.40 & 0.14 & 0.15 & 0.18 & -1.20 & 0.37 & 1.95 & 0.21 & 0.141 \\
		155 &       D631-7 & 0.99 & 0.02 & 0.96 & 0.05 & 0.39 & 0.08 & \multicolumn{2}{c}{......} & 1.80 & 0.08 & 1.11 & 0.03 & 0.83 & 0.10 & -1.94 & 0.08 & 1.89 & 0.06 & 1.001 \\
		156 &      NGC3109 & 1.00 & 0.05 & 1.00 & 0.07 & 0.50 & 0.12 & \multicolumn{2}{c}{......} & 1.75 & 0.02 & 1.22 & 0.03 & 0.67 & 0.04 & -1.67 & 0.07 & 2.00 & 0.04 & 0.214 \\
		157 &      UGCA281 & 1.00 & 0.05 & 1.00 & 0.04 & 0.50 & 0.17 & \multicolumn{2}{c}{......} & 1.20 & 0.31 & 1.58 & 0.10 & -0.25 & 0.39 & -0.72 & 0.25 & 2.03 & 0.20 & 0.570 \\
		158 &       DDO168 & 0.97 & 0.05 & 0.87 & 0.09 & 0.46 & 0.10 & \multicolumn{2}{c}{......} & 1.71 & 0.04 & 1.32 & 0.04 & 0.53 & 0.07 & -1.41 & 0.12 & 2.12 & 0.08 & 7.568 \\
		159 &       DDO064 & 1.04 & 0.30 & 1.02 & 0.08 & 0.49 & 0.13 & \multicolumn{2}{c}{......} & 1.52 & 0.32 & 1.37 & 0.15 & 0.28 & 0.42 & -1.27 & 0.39 & 2.01 & 0.28 & 0.509 \\
		160 &     PGC51017 & 1.02 & 0.09 & 1.01 & 0.04 & 0.56 & 0.10 & \multicolumn{2}{c}{......} & 2.61 & 0.41 & 0.00 & 0.20 & 2.75 & 0.51 & -4.11 & 0.37 & 1.64 & 0.39 & ... \\
		161 &      UGCA442 & 0.99 & 0.05 & 1.00 & 0.11 & 0.50 & 0.12 & \multicolumn{2}{c}{......} & 1.60 & 0.03 & 1.29 & 0.05 & 0.45 & 0.05 & -1.49 & 0.12 & 1.96 & 0.08 & 1.867 \\
		162 &     UGC07866 & 1.00 & 0.05 & 1.00 & 0.12 & 0.48 & 0.14 & \multicolumn{2}{c}{......} & 1.25 & 0.55 & 1.39 & 0.18 & -0.01 & 0.70 & -1.23 & 0.46 & 1.77 & 0.37 & 0.348 \\
		163 &     UGC07232 & 1.00 & 0.06 & 1.00 & 0.08 & 0.50 & 0.11 & \multicolumn{2}{c}{......} & 2.49 & 0.32 & 1.51 & 0.05 & 1.11 & 0.34 & -0.90 & 0.14 & 3.21 & 0.30 & ... \\
		164 &     UGC07559 & 0.99 & 0.05 & 0.99 & 0.05 & 0.49 & 0.13 & \multicolumn{2}{c}{......} & 1.39 & 0.45 & 1.22 & 0.09 & 0.31 & 0.52 & -1.66 & 0.22 & 1.65 & 0.37 & 0.521 \\
		165 &      NGC6789 & 0.99 & 0.05 & 1.01 & 0.16 & 0.50 & 0.12 & \multicolumn{2}{c}{......} & 1.75 & 0.34 & 1.75 & 0.07 & 0.13 & 0.37 & -0.27 & 0.19 & 2.87 & 0.32 & ... \\
		166 &     KK98-251 & 1.12 & 0.32 & 1.03 & 0.09 & 0.48 & 0.13 & \multicolumn{2}{c}{......} & 1.52 & 0.43 & 1.06 & 0.20 & 0.60 & 0.58 & -2.07 & 0.48 & 1.52 & 0.32 & 0.398 \\
		167 &     UGC05764 & 0.97 & 0.26 & 1.00 & 0.16 & 0.48 & 0.12 & \multicolumn{2}{c}{......} & 1.47 & 0.05 & 1.62 & 0.09 & -0.02 & 0.11 & -0.62 & 0.24 & 2.36 & 0.15 & 5.065 \\
		168 &         CamB & 0.91 & 0.08 & 0.90 & 0.08 & 0.34 & 0.07 & \multicolumn{2}{c}{......} & 2.67 & 0.27 & 0.97 & 0.06 & 1.84 & 0.28 & -2.29 & 0.15 & 2.55 & 0.28 & 4.135 \\
		169 &  ESO444-G084 & 1.01 & 0.10 & 1.04 & 0.22 & 0.55 & 0.14 & \multicolumn{2}{c}{......} & 1.57 & 0.08 & 1.50 & 0.13 & 0.20 & 0.15 & -0.94 & 0.33 & 2.27 & 0.22 & 7.899 \\
		170 &       DDO154 & 1.00 & 0.05 & 1.00 & 0.05 & 0.43 & 0.08 & \multicolumn{2}{c}{......} & 1.54 & 0.01 & 1.26 & 0.02 & 0.42 & 0.03 & -1.56 & 0.06 & 1.86 & 0.04 & 1.896 \\
		171 &     UGC07577 & 1.00 & 0.05 & 1.00 & 0.05 & 0.49 & 0.11 & \multicolumn{2}{c}{......} & 2.51 & 0.37 & 0.82 & 0.20 & 1.83 & 0.44 & -2.65 & 0.42 & 2.18 & 0.41 & 0.296 \\
		172 &       D564-8 & 1.00 & 0.03 & 1.00 & 0.11 & 0.51 & 0.13 & \multicolumn{2}{c}{......} & 1.33 & 0.34 & 1.14 & 0.09 & 0.33 & 0.41 & -1.87 & 0.22 & 1.46 & 0.25 & ... \\
		173 &      NGC3741 & 1.03 & 0.05 & 1.03 & 0.05 & 0.77 & 0.17 & \multicolumn{2}{c}{......} & 1.56 & 0.02 & 1.23 & 0.03 & 0.47 & 0.04 & -1.65 & 0.08 & 1.82 & 0.04 & 0.988 \\
		174 &     UGC04483 & 0.99 & 0.10 & 1.00 & 0.05 & 0.49 & 0.15 & \multicolumn{2}{c}{......} & 1.11 & 0.33 & 1.52 & 0.12 & -0.27 & 0.44 & -0.89 & 0.32 & 1.84 & 0.20 & 0.744 \\
		175 &      UGCA444 & 1.00 & 0.05 & 0.99 & 0.05 & 0.51 & 0.12 & \multicolumn{2}{c}{......} & 1.33 & 0.05 & 1.38 & 0.04 & 0.09 & 0.09 & -1.25 & 0.12 & 1.83 & 0.05 & 0.245 \\
		\hline
	\end{longtable}
\end{landscape}

\vfill
\end{document}